\documentclass[a4paper, 11pt]{article} 
\usepackage[margin=2.5cm, top=2.5cm, bottom=3cm, footskip=1cm]{geometry}	

\usepackage{amsmath}		
\usepackage{amsfonts}		
\usepackage{amsthm}
\usepackage[utf8]{inputenc}
\usepackage{color}
\usepackage{tikz}
\usepackage{slashed}
\usepackage{mathrsfs} 
\usetikzlibrary{calc}
\usetikzlibrary{arrows}  
\usetikzlibrary{decorations.markings}
\usetikzlibrary{shapes,snakes}
\usepackage{hyperref} 
\hypersetup{
	bookmarks=true,
	unicode=false,
	pdftoolbar=true,
	pdfmenubar=true,
	pdffitwindow=false,
	pdfstartview={FitH},
	pdftitle={My title},
	pdfauthor={Author},
	pdfsubject={Subject},
	pdfcreator={Creator},
	pdfproducer={Producer},
	pdfkeywords={keyword1} {key2} {key3},
	pdfnewwindow=true,
	colorlinks=true,   
	linkcolor={blue!80!black}, 
	citecolor={CrimsonRed}, 
	filecolor={blue!80!black},
	urlcolor={blue},      
	hypertexnames=false, %
} 
\usepackage[figure]{hypcap}
\usepackage{amssymb}		
\usepackage{amstext}		
\usepackage{graphicx}		
\usepackage[utf8]{inputenc}	
\usepackage[T1]{fontenc}	
\usepackage{pdfpages}		
\usepackage{wrapfig}		
\usepackage{here}			
\usepackage{float}			
\usepackage{fancyhdr}
\usepackage{verbatim}		
\usepackage{wasysym}		
\usepackage{tocloft}		
\usepackage{siunitx}		
\usepackage[arrow, matrix, curve]{xy} 
\usepackage{wrapfig,tabularx,multirow,hyperref,icomma}
\usepackage{mathptmx}
\usepackage{lmodern}
\usepackage{courier}
\usepackage{paralist}
\usepackage{afterpage}
\usepackage{subcaption}
\usepackage[section]{placeins}
{\left.\begin{aligned}}
	{\end{aligned}\right\rbrace}
	
\usepackage{etoolbox}
\usepackage{relsize,exscale}
\usepackage[titletoc]{appendix}
\usepackage{natbib}
\usepackage[bb=boondox]{mathalpha}

\usepackage{nameref}
\definecolor{RoyalBlue}{RGB}{35,32, 57}
\definecolor{lightblue}{RGB}{132,131,143}
\definecolor{CrimsonRed}{RGB}{130,0,29}

\theoremstyle{definition}

\theoremstyle{definition}

\theoremstyle{definition}

\theoremstyle{definition}

\renewcommand{\Im}{\mathrm{Im}} 
\renewcommand{\Re}{\mathrm{Re}} 

\numberwithin{equation}{section}

\makeatletter
\patchcmd{\chapter}{\if@openright\cleardoublepage\else\clearpage\fi}{}{}{}
\makeatother

\usepackage{graphicx}
\usepackage{calc}
\usepackage{tocloft}
\usepackage[nottoc,notlof]{tocbibind}
\usepackage{emptypage}
\usepackage{hyperref}
\usepackage{bashful}
\usepackage{enumerate}

\usepackage{authblk}

\title{Extension of a mathematical analogy between vortical flows without surface
	tension and irrotational capillary flows}

\author[1]{Valentin D. Kunz}
\affil[1]{{\footnotesize Department of Mathematics, Ohio State University, 231 W 18th Ave, Columbus, OH 43210, USA}}
\author[1]{Saleh  Tanveer}
\author[2]{Darren G. Crowdy}
\affil[2]{{\footnotesize Department of Mathematics, Imperial College London, 180 Queen's Gate, London SW7 2AZ, UK }}
\affil[ ]{{\footnotesize
		v$\_$kunz@startmail.com	}}
\date{\today}
\begin{document}

\maketitle

\begin{abstract}
	In a recent paper, a mathematical analogy was found between 
	two physically distinct classes of  two-dimensional free boundary problems in inviscid fluid mechanics. One class involves
	free surface flows with constant vorticity in the bulk but without boundary surface tension, the other
	comprises free surface flows without vorticity in the bulk but with non-zero boundary surface tension (capillarity).
	Several examples 
	were given
	of pairs of physically-distinct free boundary problems that amount to solving the same mathematical problem, albeit
	with the possibility of a single constraint arising for the capillary flows.
	The present work shows that
	this analogy extends to
	another pair of physically-distinct problems:
	a  hollow vortex equilibrium in a shear flow with a third
	order irrotational strain
	and a steady bubble with surface tension in a simple straining flow.
	For both problems, we construct an isolated exact solution when a parameter $\gamma=0$ 
	and provide numerical solutions for non-zero $\gamma$. 
	An interesting feature is that non-trivial conformal mapping identities
	emerge from the class of problems with capillarity 
	that apply equally for the hollow vortex problems when
	the parameters are appropriately constrained.
\end{abstract}
\normalsize

\maketitle
	\section{Introduction}
	Recent work has shown that several physically-distinct free boundary problems in inviscid fluid dynamics admit the same free-surface profiles as solutions. One case arises in the study of water waves. Hur \& Wheeler \cite{Miles} 
have   shown analytically that the shapes of steadily-translating 
water 
waves, without
surface tension,
on a uniform shear current can adopt exactly the ``Crapper-wave'' profiles
famous in fluid mechanics as being the shapes of irrotational pure-capillary waves
on deep water \cite{Crapper}. 
Mounting numerical evidence  \cite{Vanden2, Hur2, Hur3} had pointed to this
being the case motivating the investigation by Hur \& Wheeler \cite{Miles}.
In the first problem, vorticity is present in the bulk but there is no surface tension
on the boundary; in the capillary-wave problem, there is no vorticity in the bulk
but surface tension is present on the wave surface.

On a different front, Crowdy {\em et al} \cite{Hstate} recently derived exact solutions
for the conformal mappings describing
the shapes of a steadily-rotating hollow vortex without surface tension, to which the authors gave the name ``H-states'',
and found them to have the same functional form as those found much
earlier by Crowdy \cite{CroCirc} and Wegmann \& Crowdy \cite{Wegmann}
describing the shapes of a non-rotating hollow vortex with uniform surface
tension active on its boundary.

Motivated by this recurring circumstance of two physically distinct free boundary
problems sharing mathematical
solutions,
Crowdy \& Tanveer \cite{CrowdyTanveer2025} 
have furnished a mathematical rationalization for these observations.
They pointed out that 
a common mathematical framework accommodates both classes of problems.
They showed that both the rotating H-state solutions of 
Crowdy {\em et al} \cite{Hstate} and the non-rotating hollow vortices
with surface tension of Crowdy \cite{CroCirc} and Wegmann \& Crowdy \cite{Wegmann}
are solutions of what they called ``Problem 1'' stated mathematically as follows.

\subsection{Statement of Problem 1 \label{Problem1}}

{\it Determine a simple closed curve $z={\mathcal Z}(s)$ that satisfies 
	\begin{equation}
		\label{1.10a}
		{d {\mathcal Z}(s) \over ds} = -{\rm i} \gamma {\mathcal Z}(s) + {\bar \xi} \left ( \overline{\mathcal Z} (s) \right ),
	\end{equation}
	for some {\it a priori} unknown 
	analytic function $\xi(z)$ in 
	the domain $\Omega$ exterior of $z={\mathcal Z} (s)$
	with
	\begin{equation}
		\label{eq1.problem1a}
		\xi (z) \sim  O(1/z), \qquad {\rm as } ~z \rightarrow \infty,
	\end{equation}
	where $\gamma \in \mathbb{R}$ is specified.
}

In this case, the solutions to Problem 1, and indeed to both physical problems, are given by 
conformal mappings that are
rational functions.

Crowdy \& Tanveer \cite{CrowdyTanveer2025} 
went on to show that another 
pair of generalized problems are also analogous in the same fashion, but with a caveat.
They showed that
the problem of a hollow vortex in a simple shear and a linear strain
and the problem of a steadily-translating bubble with surface tension on its boundary 
have solutions within the class of solutions of ``Problem 2'' stated as follows.

\subsection{Statement of Problem 2 \label{Problem2}}

{\it Determine a simple closed curve $z={\mathcal Z}(s)$ that satisfies 
	\begin{equation}
		\label{1.10}
		{d {\mathcal Z} \over ds} = -{\rm i} \gamma {\mathcal Z} + {\bar \xi} \left ( \overline{\mathcal Z} (s) \right ) ,
	\end{equation}
	for some {\it a priori} unknown 
	analytic function $\xi(z)$ in 
	the domain $\Omega$ exterior of $z={\mathcal Z} (s)$, except at $\infty$, where
	\begin{equation}
		\label{eq1.problem1}
		\xi (z) \sim - {\rm i} b z + O(1) \ , {\rm as } ~z \rightarrow \infty 
		\ ,
	\end{equation}
	and where $(\gamma, b)$ is specified.
}

It is clear that Problems 1 and 2 differ only in the singularity at infinity of the function 
$\xi (z)$: in Problem 1, $\xi(z)$ is not singular there, and in Problem 2 it has a simple pole.
The caveat is that, for the bubble problem, only solutions
of Problem 2 that satisfy a single additional constraint 
between $b$ and $\gamma$ are physically admissible
solutions. This single constraint leads to a surprising, non-trivial, conformal
mapping identity that holds for any solution of Problem 2 in the presence of this constraint.

  {The principal purpose} of the present paper is to
extend this analogy further and
show that there is yet another pair of physically important free boundary
problems -- to be called Problem (a) and Problem (b) and detailed below -- that are also analogous in the same sense as for Problem 2 above. That is, both physical problems 
admit solutions that
are solutions of a new  mathematical problem, ``Problem 3', which can be stated as follows.

\begin{center}
	\begin{figure}
		\includegraphics[scale=0.35]{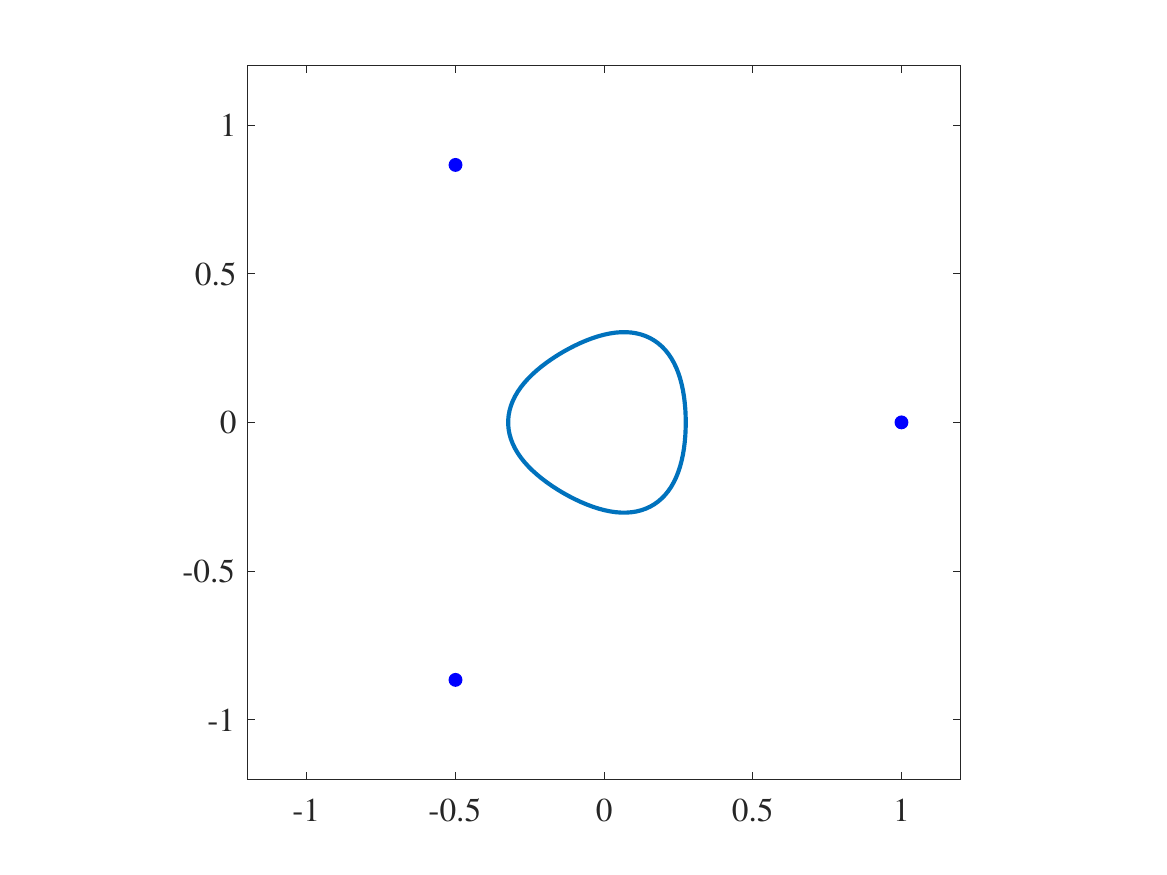}
		\includegraphics[scale=0.35]{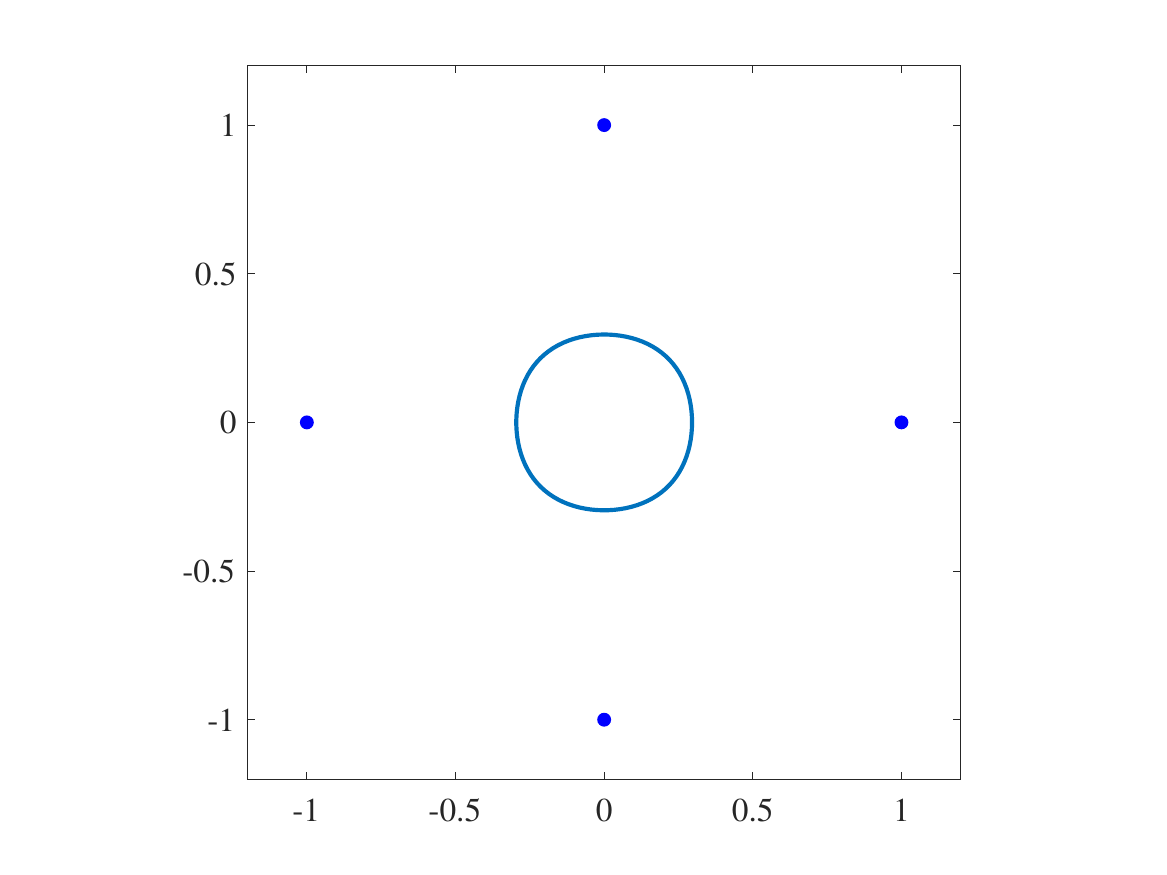}
		\caption{Typical solutions of Crowdy \& Roenby \cite{Roenby}
			for a central hollow vortex of circulation $\Gamma_0$
			with $N = 3$ (left) and $N=4$ (right) satellite point vortices each of circulation $\Gamma$.
			The value of $\Gamma_0/\Gamma$ is such that the configuration is a non-rotating equilibrium.
			\label{CRFig}}
	\end{figure}
\end{center}

\subsection{Statement of Problem 3 \label{Problem3}}

{\it Determine a simple closed curve $z={\mathcal Z}(s)$ that satisfies 
	\begin{equation}
		\label{1.10}
		{d {\mathcal Z} \over ds} = -{\rm i} \gamma {\mathcal Z} + {\bar \xi} \left ( \overline{\mathcal Z} (s) \right ) ,
	\end{equation}
	for some {\it a priori} unknown 
	analytic function $\xi(z)$ in 
	the domain $\Omega$ exterior of $z={\mathcal Z} (s)$, except at $\infty$, where
	\begin{equation}
		\label{eq1.problem1}
		\xi (z) \sim  -{\rm i}  ( \lambda z^3  +{\bar \sigma} z  )  
		+ O(1) 
		\qquad {\rm as } ~z \rightarrow \infty 
		\ ,
	\end{equation}
	and where $(\lambda, {\bar \sigma}, \gamma)$ is specified.
}

Evidently, the difference is that $\xi(z)$ now has a third-order pole at infinity. 
The structure of the analogy is, however, identical to that associated with Problem 2:
the solutions of
both Problem (a) and (b) fall within the class of solutions of Problem 3 with the caveat that,
for the bubble problem, only solutions
of Problem 3 that satisfy a single additional constraint between parameters are physically admissible
solutions. Once again this single constraint leads to another surprising, non-trivial, conformal
mapping identity that holds for any solution of Problem 3 in the presence of this constraint.

Before describing Problems (a) and (b) 
that are the focus of this study,
it is worth explaining
how this hierarchy of higher-order problems is relevant to applications.

Although it is an old model \cite{Pocklington1895},
research
interest in the hollow vortex model has experienced an explosion
in recent years, not least because of its amenability
for modelling compressible flows
\cite{baker-hv, Ardalan, CG:VKSHV, Compress, crowdy-thv, Roenby, Zannetti, Recent, Hstate}.
Among the many recent contributions to the literature,
Crowdy \& Roenby \cite{Roenby} have found analytical solutions for the shape and circulation of
a central hollow vortex of circulation $\Gamma_0$
surrounded by $N \ge 2$ point vortices, all of circulation $\Gamma$, at the vertices
of a regular $N$-gon with $\Gamma_0/\Gamma$ chosen
such that the
entire vortical structure is in equilibrium, i.e., not rotating. Typical solutions for $N=3$ and $4$ are shown
in Figure \ref{CRFig}; these are the images of the unit disc under a rational function
conformal mapping derived by Crowdy \& Roenby \cite{Roenby}.
It is natural to ask:
for values of $\Gamma_0/\Gamma$ not equal to this equilibrium
value,
can a modified relative equilibrium be found
with $N$-fold rotational symmetry that is now steadily rotating?
Such rotating configurations are certainly
possible when the central vortex is a point vortex, in which case
the results are classical \cite{MS, saffman-vdynamics}, 
and when the central vortex is modelled as a vortex patch \cite{CroJFM}. It therefore seems likely
that analogous steadily-rotating solutions will exist if the central vortex is hollow.
Such a configuration will be steady in a frame of reference co-rotating with the angular
velocity of the structure, but the flow will now have an additional uniform vorticity
component associated with the move to the corotating frame.
A first step in investigating the existence of such relative equilibria would be
to carry out an analysis using matched asymptotics when the effective
radius of the central hollow vortex
is small compared to the distance  from its centroid of the point vortices.
In that case, 
an inner problem close to the hollow vortex would have to match, as the inner complex
variable $z \to \infty$,
to a straining flow of degree $N-1$ where $u-{\rm i}v$ is of order $z^{N-1}$.
For $N=2$ this inner problem is 
akin to that 
of a hollow vortex in simple shear
as considered by  Zannetti {\em et al} \cite{Zannetti} and recently
reappraised, for a more general shear and and linear far-field strain,
by Crowdy \& Tanveer \cite{CrowdyTanveer2025} where it was
shown to have the same mathematical structure as a completely different physical problem: that of
a steadily-translating bubble with uniform surface tension active on its boundary.

Consider, however, seeking the steadily-rotating analogue of the $N=4$ solution
of Crowdy \& Roenby \cite{Roenby} shown in Figure \ref{CRFig}.
The inner problem in this case
is that of determining the shape of
a hollow vortex
in the same linear shear and and linear far-field strain but with an additional
third-order far-field irrotational straining flow component. 
The problem we consider
here is more general since no relation between linear strain component and shear is required. 
Then,
in view of the analogy elucidated 
by Crowdy \& Tanveer \cite{CrowdyTanveer2025},
it is natural to ask if this generalized hollow vortex problem has
an ``analogous'' bubble problem with surface tension.
These considerations have motivated 
the present paper. 

The answer is
in the affirmative and, intriguingly, the analogous capillary bubble problem is one
that has already received attention in the literature: the problem of a bubble
with uniform surface tension sitting in an ambient linear
straining flow. Vanden-Broeck \& Keller \cite{Vanden} carried out a numerical investigation
of this flow scenario and found a class of equilibrium bubble shapes when there is no circulation
around the bubble.
Therefore, the purpose of the present paper is to point out and examine
a mathematical analogy 
between:

{\bf Problem (a):}
an equilibrium hollow vortex with no surface tension 
in a third order straining flow, also allowing for a first-order strain component
and
a simple shear. For convenience, 
since this flow has uniform vorticity,
this will be referred to here as the ``vortex problem''.

{\bf Problem (b):} a bubble, possibly with circulation around it, with surface tension 
and situated
in a first order straining flow.
For convenience, since surface tension is the key feature here,
this will be referred to as the ``bubble problem'' (even though
there may also be circulation around the bubble making it
more akin to a hollow vortex with surface tension).

The analogy  between these two physically-distinct problems
can be understood as a ``higher-order'' extension of the analogy already
studied by Crowdy \& Tanveer \cite{CrowdyTanveer2025} 
for a different pair of mathematically-equivalent flows with
lower-order far-field flow behaviour.
Both have solutions that are solutions of Problem 3 given above. 
However, 
not all solutions of Problem 3 are relevant to the bubble problem; there is a constraint
between parameters $(\gamma, \lambda, \sigma)$ that needs to be satisfied as has been proved in the
companion paper \cite{Valentin}.

\begin{figure}
	\begin{center}
		\includegraphics[scale=0.5]{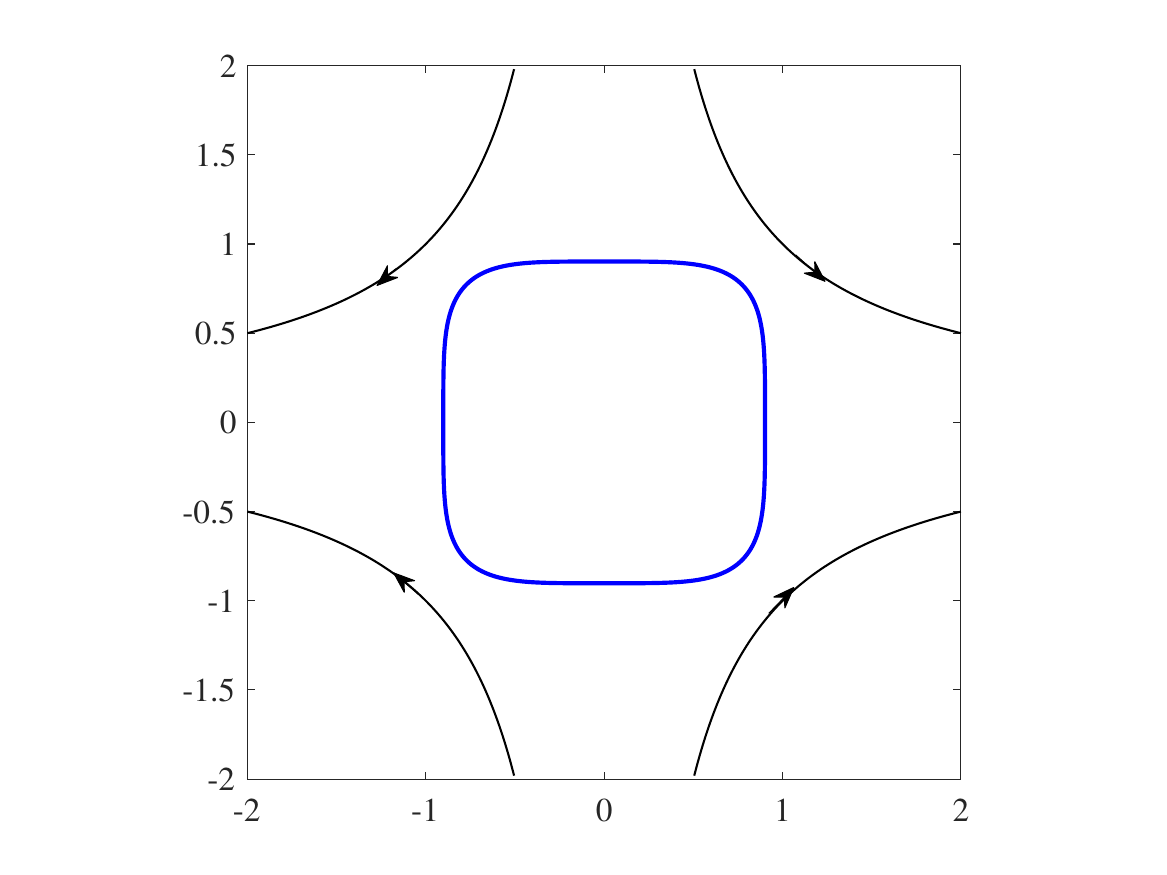}
		\caption{Bubble shape given by the exact solution (\ref{newexact}) for a bubble, with surface tension and zero circulation, in a linear
			straining flow. This isolated exact solution for vanishing Bernoulli constant $\gamma=0$ is the analogue,
			for far-field strain conditions,
			of the classical McLeod \cite{Mcleod}  exact solution for a uniformly-travelling bubble.
			\label{VKFig}}
	\end{center}
\end{figure}

With this analogy set out, another objective
of the present paper is to elucidate the physical features of the solutions to these two problems.
This includes identifying exact solutions when a parameter $\gamma=0$: this corresponds to
no shear/rotation in the vortex problem, and 
to a vanishing Bernoulli constant
in the bubble problem.
These solutions are extended numerically for non-zero $\gamma$.
It is worth pointing out that, for the bubble in a linear strain,
it is found here that when $\gamma=0$ and there is no circulation around the bubble,
the steady bubble shape is given by
the rational-function conformal mapping
\begin{equation}
	z = Z(\zeta) = - {1 \over \zeta} 
	{+} {2 \zeta^3 \over 21} + {\zeta^7 \over 343}.
	\label{newexact}
\end{equation}
The existence of this explicit solution for $\gamma=0$
does not
appear to have
been recognized by Vanden-Broeck \&
Keller\cite{Vanden} 
who numerically calculated solutions for zero-circulation bubbles at arbitrary $\gamma$.
The exact solution (\ref{newexact}), a new result of this paper, can be thought of as
being
analogous to a similar isolated exact solution
for a  steadily-translating bubble found by
McLeod \cite{Mcleod}, also in the 
situation when the Bernoulli constant vanishes. 
  {Exact solutions for bubbles also extend to 
	nonzero circulation $\Gamma_0$ (presented later in 
	a unified vortex-bubble approach) in  
	\eqref{e.3} with parameters $(\lambda, \sigma)= 
	(\alpha^2/3, 2 {\rm i} \alpha \Gamma_0  ) $ 
	constrained by \eqref{b.11.1} and the univalence of the resulting 
	$Z(\zeta)$.}

For general non-zero values of $\gamma$, 
although no analytical solutions
are known, Tanveer \cite{T96} has
shown that the square root of the derivative of a conformal mapping from a unit disc relevant to this problem
admits a meromorphic analytic continuation outside the disc.  
  {While Tanveer \cite{T96} concerned a translating bubble,
	a variant of the argument used  
	applies to the bubble in 
	a straining flow.}

In a companion paper \cite{Valentin}, the existence of solutions 
for small $\gamma$ is proven, including proof of the univalence of the mapping function,
and that an apparently overdetermined system of equations
in the case of the bubble problem is, despite appearances, consistent even for $\gamma \ne 0$. 
This automatic consistency
leads to a conformal mapping identity, akin to one already found
by Crowdy \& Tanveer \cite{CrowdyTanveer2025},
that transcends the specifics of the bubble problem and holds equally for solutions of the
hollow vortex problem when 
a constraint is satisfied between the first, third order components  
of strain and the shear rate.

\section{Hollow vortex with uniform vorticity and third-order strain}

Since the hollow vortex problem has been set out carefully
by Crowdy \& Tanveer \cite{CrowdyTanveer2025}, and given that the only difference here
is the nature of the far-field conditions, only a brief description of the governing
physical problem will be given here.

Consider 
two-dimensional flow of an ideal, inviscid fluid.
It is assumed that
a single hollow vortex is situated in a simple shear 
and a third-order irrotational straining flow, allowing also for an additional linear irrotational strain component. 
(The ambient uniform vorticity associated with the simple shear may have its origins in moving
to a frame of reference corotating with some angular velocity, as would arise in the motivating
problem described above.)
Let ${\bf u} = (u,v)$ be the velocity in the fluid domain exterior of
the hollow vortex.  
By the boundary conditions
that, for a steady solution, the boundary of the hollow vortex must be a streamline
and given that a form of the Bernoulli theorem pertains \cite{saffman-vdynamics},
it can be shown that the fluid
speed on the hollow vortex boundary is constant, and can be taken to be unity by a choice of the velocity scale.
Then, with a counter-clockwise boundary flow 
\begin{equation}
	\label{h.1}
	u-{\rm i}  v = 
	\frac{d}{ds} \overline{\mathcal{Z}} (s)
\end{equation}
on $z = \mathcal{Z} (s)$, where $s$ is the arclength increasing counter-clockwise.
Furthermore,  for $z \in \Omega $ exterior of the simple closed curve $\mathcal{Z} (s)$,
\begin{equation}
	\label{h.2}
	u-{\rm i} v = {\rm i} {\gamma} {\overline z} + \xi (z) \ ,   
\end{equation}
where $2 \gamma$ is the constant shear rate and $\xi$ is analytic in $\Omega$. 
The bar symbol ${\overline {\cdot}} $ appearing in 
\eqref{h.2}, and in what follows, denote complex conjugation when applied
to a variable and the conjugate  
function when applied to an analytic function.\footnote{${\bar f} (z)$ 
	is the analytic function conjugate to $f(z)$ if
	for $z=x\in \mathbb{R}$,
	${\bar f} (x) = \overline{f(x)}$. 
	Note ${\bar f} (z) \ne \overline{f (z)}$ for $z \notin \mathbb{R}$.}
Far-field conditions on the velocity imply
\begin{equation}
	\label{h.3}
	\xi (z) \sim  {\rm i} \left ( {\hat \alpha} z^3  + (\beta -\gamma ) z  
	+ \delta +\omega z^{-1} \right ) + O (z^{-2} ) 
	\ .
\end{equation}
The parameter $\gamma $ is taken to be real corresponding to a specific orientation of the shear.  
The circulation parameter $ \omega$ must be real, 
otherwise there will be a source/sink
at infinity which 
is incompatible with a steady vortex with unchanging area. 
It will turn out that, all within a certain range,
the third and first order components of strain 
${\hat \alpha}$ and $\beta$, respectively, as well as 
$\gamma$, the shear rate, can be specified.
The drift
$\delta$, 
and circulation\footnote{The actual circulation is $-2 \pi \omega$} $\omega$ 
are determined as part of the solution.  

From \eqref{h.1} and \eqref{h.2},  
on the interface $z=\mathcal{Z} (s)$, 
\begin{equation}
	\label{h.4}
	\frac{d}{ds} \overline{\mathcal{Z}} (s) = {\rm i}  {\gamma} 
	\overline{\mathcal{Z} (s)} + \xi \left ( \mathcal{Z} (s) \right ) 
\end{equation}
The parameter $\hat \alpha$ can be chosen, without loss of generality, to satisfy ${\hat \alpha} <  0$.\footnote{We are grateful to Ali Farsad
	for pointing out this symmetry in the context of a vortex in a simple strain and shear}
To see this,
note that
on setting $z=e^{{\rm i}  \phi} {\hat z}$, $\mathcal{Z} (s) = e^{{\rm i}  \phi} \mathcal{\hat Z} (s)$ and
$\xi (z) = e^{-{\rm i}  \phi} {\hat \xi} \left ( e^{-{\rm i}  \phi} z \right ) $, then equation
\eqref{h.4} reproduces itself in the hatted variables 
with the asymptotic condition (1.3) 
translating to 
\begin{equation}
	\label{h.4.1}
	{\hat \xi} ({\hat z}) \sim {\rm i}  \left ( e^{-4 {\rm i}  \phi} {\hat \alpha} {\hat z}^3  
	+ e^{-2{\rm i}  \phi} (\beta{-\gamma}) {\hat z} + e^{-{\rm i}  \phi} \delta + \omega {\hat z}^{-1} \right ) + O \left ({\hat z}^{-2} \right ) .
\end{equation}
Therefore, $\phi$ can be chosen so that $e^{-4{\rm i}  \phi} {\hat \alpha} 
\in \mathbb{R}^-$.

The interface is described by introducing
the conformal map $Z: \mathbb{D} \rightarrow \Omega$ with 
$\zeta=0$ corresponding to $\infty$: 
\begin{equation}
	\label{h.5}
	z=Z (\zeta) = - \zeta^{-1} + \sum_{j=2}^\infty c_j \zeta^{j-1}.
\end{equation}
for some set of coefficients $\lbrace c_j | j \ge 2 \rbrace$.
The rotational degree of freedom in the Riemann mapping theorem has been
utilized by demanding the coefficient of $\zeta^{-1}$ to be negative while
the length scale of $Z$ has been set 
implicitly
by demanding the coefficient of $-\zeta^{-1}$ to
be unity. 
The absence of the $j=1$ term in the summation
in \eqref{h.5} reflects a  
a choice of origin in the $z$-plane.
An important observation is that
\begin{equation}
	y(\zeta) \equiv \left( {d Z \over d\zeta} \right)^{\frac{1}{2}}
	\label{yeqn}
\end{equation}
has the convergent representation in $\mathbb{D}$:
\begin{equation}
	\label{h.6}
	y(\zeta) = \zeta^{-1} + \sum_{j=2}^\infty a_j \zeta^{j-1},
\end{equation}
where $\left \{ a_j \right \}_{j=2}^\infty$ and $\left \{ c_j \right \}_{j=2}^\infty$ are related through
\begin{equation}
	\label{h.7}
	c_j = \frac{2 a_j}{j-1}  + \frac{1}{j-1} \sum_{m=2}^{j-2}  a_m a_{j-m}  \  , \quad {\rm for} ~j \ge 2
\end{equation}
and
where the summation term 
is to be understood to be zero when
$j < 4$.
It will be convenient 
to use the abbreviated notation
\begin{equation}
	\label{h7.1}
	Z(\zeta) = \sum_{j=0}^\infty c_j \zeta^{j-1} \ , \quad y(\zeta) = 
	\sum_{j=0}^\infty a_j \zeta^{j-1}  \ , \quad {\rm where}~
	c_1=0=a_1, \quad c_0=-1, a_0=1.
\end{equation}
Another analytic function
$\overline{K} (\zeta)$ is defined in $\mathbb{D}\setminus \{0\}$ 
by
\begin{equation}
	\label{h.8}
	\xi \left ( Z (\zeta) \right ) = {\rm i}  \overline{K} (\zeta).
\end{equation}
Since the velocity has to be single valued in the flow domain $\Omega$ 
so must $\overline{K} (\zeta)$ in
$\mathbb{D} \setminus \{0 \}$.
Since $\xi (z)$ is analytic in $\Omega$ with asymptotic behavior
\eqref{h.3}, the transformation in $\zeta$ through \eqref{h.5}, \eqref{h.8}
gives a convergent Laurent expansion in $\mathbb{D}$:
\begin{equation}
	\label{h.9}
	\overline{K} (\zeta) = \sum_{j=-2}^\infty {\bar k}_j \zeta^{j-1}  
	\ , \quad {\rm where} ~{\bar k}_{-1} = 0 
\end{equation}    
where the first few terms are
\begin{equation}
	\label{h.10}
	{\bar k}_{-2} = - {\hat \alpha} \ , \quad {\bar k}_{0} =  
	3 {\hat \alpha} c_2 -\beta{+\gamma}  \ , \quad {\bar k}_1 = 
	3 c_3 {\hat \alpha} + \delta \ , \quad 
	{\bar k}_2 = 3 {\hat \alpha} c_4 - 3 c_2^2 {\hat \alpha} + (\beta-{\gamma}) c_2 -\omega.
\end{equation}
Since 
$\mathcal{Z}_s = -{\rm i} \zeta y (\zeta)/{\bar y} (\zeta^{-1}) $, 
complex conjugation of
\eqref{h.4} on $|\zeta|=1$  leads to 
\begin{equation}
	\label{h.11}
	\frac{\zeta y (\zeta)}{{\bar y} \left (\zeta^{-1}\right )}  - \gamma 
	Z (\zeta)  =  
	K(\zeta^{-1}) , 
\end{equation}
a functional equation that is equally valid for the hollow vortex
and the bubble, as will be seen in what follows. 
Note that in writing \eqref{h.11} it is implicitly assumed that $\bar{y}(\zeta^{-1})$ is non-vanishing on $|\zeta|=1$, which is indeed always true, as has been proved in \cite{Valentin}.

Though \eqref{h.11} is a single complex equation, it turns out to be enough
to determine both the shape and velocity-related functions
$Z(\zeta)$ and $K (1/\zeta)$, and hence
the conjugate function $\overline{K} (\zeta) $, for specified $\gamma$ within
a certain range, 
once the Laurent series coefficients 
$k_{-2}, k_0$ are specified. In the case of the vortex problem this is equivalent to
to specifying strain-rate components ${\hat \alpha}$ and $\beta$.

Another 
important feature of \eqref{h.11} is that it can be analytically
continued off the unit circle when the solution is analytic on
$|\zeta|=1$, a property proved in \cite{Valentin} for a solution for
which the series \eqref{h.6} is absolutely convergent on $|\zeta|=1$.
Through analytic continuation, as with
the translating bubble \cite{T96} \cite{CrowdyTanveer2025}
or hollow vortex in linear strain and shear\cite{CrowdyTanveer2025}, 
the function $y(\zeta)$ as well as $Z(\zeta)$ may be
shown to be meromorphic in $\mathbb{C}$ for $\gamma \ne 0$, although this feature will
not be discussed here in any detail. All the poles in $|\zeta| > 1$ 
approach $\infty$ as $\gamma \rightarrow 0$ and,
for $\gamma=0$, the meromorphic function
degenerates into a simple rational function with a simple pole
at the origin. The latter constitute new exact solutions
to the various free boundary problems considered here.

\section{Bubble, with surface tension, in a linear strain}

Consider 
now the different free boundary problem of a bubble in an inviscid flow 
with linear strain at infinity
and with surface tension active on its surface.
The flow is assumed to be irrotational although the possibility of
non-zero circulation around the bubble is allowed for.
This problem has been studied numerically by 
Vanden-Broeck \& Keller \cite{Vanden} in the case where there is no circulation around the
bubble. 

If ${\bf u} = (u, v)$ denotes fluid velocity, then the 
the complex velocity $u-{\rm i}v$ is
complex analytic function of $z=x_1+{\rm i} x_2$, denoted by ${dW}/{dz}$. The asymptotic
behavior of the flow-field at $\infty$ translates to
\begin{equation}
	\label{b.1}
	u-{\rm i} v = \frac{dW}{dz} \sim \alpha z - \frac{i {\Gamma_0}}{z} + O (z^{-2}),
\end{equation}
where, without any loss of generality, $\alpha > 0$. As in the hollow vortex problem, the 
circulation parameter ${\Gamma_0}$ has to be real.\footnote{Complex ${\Gamma_0}$
	implies source/sink at $\infty$ which is incompatible with a steady bubble 
	of fixed area.}
Introducing the conformal map
$Z: \mathbb{D} \rightarrow \Omega$ from the unit disk to 
the fluid flow domain $\Omega$ exterior to the simple closed curve $\mathcal{Z} (s)$,
with $\zeta=0$ corresponding to $\infty$, with scaling, rotation and $z$-origin
choice so that
$Z(\zeta)$ has the same representation
\eqref{h.5} as for the hollow vortex problem just discussed.
{
	On appropriate choice of a velocity scale, the
	Bernoulli condition on the interface may be written
	as
	\begin{equation}
		\label{b.1.0.0}
		-{\rm i} \frac{\overline{\mathcal{Z}}_{ss}}{\overline{\mathcal{Z}}_s}  
		+ \Big | \frac{dW}{dz}  \Big |^2 = \gamma, \qquad {\mathcal{Z}_s} \equiv {d {\cal Z} \over ds}, 
		\qquad  {\mathcal{Z}_{ss}} \equiv {d^2 {\cal Z} \over ds^2},
	\end{equation}
	where $s$ is the arclength oriented in the positive sense. The first
	term incorporates the surface tension, gravity has been
	ignored, and $\gamma$ is a scaled Bernoulli constant. On multiplying   
	\eqref{b.1.0.0}} by 
${\rm i} \overline{d \mathcal{Z} /ds}$ and  after integration with respect to arclength, the equation becomes
\begin{equation}
	\label{b.1.0}
	\frac{d}{ds} \overline{\mathcal{Z}}  = {\rm i} \gamma \overline{\mathcal{Z}} - {\rm i}
	\int \left | \frac{dW}{dz} \right |^2 \overline{\mathcal{Z}}_s ds 
	\ . 
\end{equation}
Use of the streamline condition in the form $\overline{\frac{dW}{dz}} 
\overline{dz} = \frac{dW}{dz} dz $ on the interface 
reduces \eqref{b.1.0} to 
\begin{equation}
	\label{b.1.1}
	\frac{d}{ds} \overline{\mathcal{Z}}  = 
	{\rm i} \gamma \overline{\mathcal{Z}} + \xi \left ( \mathcal{Z} (s) \right )
	\ ,
\end{equation}
where
\begin{equation}
	\label{b.1.1.0}
	\xi (z) = -{\rm i}
	\int^z \left ( \frac{dW}{d\tilde z} \right )^2 d\tilde z 
	\ .
\end{equation}
Equation \eqref{b.1.1} has the same mathematical form
as 
\eqref{h.4}, except that, in this bubble problem, 
there is the 
ancillary requirement that $\xi (z)$ is given by
\eqref{b.1.1.0}.
As before, the function $\overline{K}(\zeta)$ is introduced, analytic in 
$\mathbb{D} \setminus \{0 \}$, given by 
\begin{equation}
	\label{b.1.2}
	\xi \left ( Z (\zeta) \right ) = {\rm i} \overline{K}(\zeta) .  
\end{equation}
As for the 
hollow vortex,
\eqref{b.1.1} and \eqref{b.1.2} leads to
\eqref{h.11}. 
However, unlike in the vortex problem, the bubble problem comes with constraints, to be explained next,
associated with \eqref{b.1.1.0}.

The streamline condition on $\partial \Omega$, and the far-field flow requirements, imply
that the complex velocity potential\footnote{Recall that the flow in the bubble
	problem is 
	irrotational} is given, as a function of $\zeta$, by 
\begin{equation}
	\label{b.2}
	W (Z(\zeta)) ={\alpha \over 2} \left ( \zeta^2 + \zeta^{-2} \right ) + 
	{\rm i} {\Gamma_0} \log \zeta.
\end{equation}
It is easy to check that $\Im W =0$ on $|\zeta|=1$ 
while the far-field condition \eqref{b.1}, which implies
$ W(z) \sim \tfrac12 \alpha z^2 - {\rm i} {\Gamma_0} \log z + o (1/z)$,
is satisfied by $W (Z(\zeta))$ as $\zeta \rightarrow 0$ due
to \eqref{h.5}.
The relation 
\eqref{b.1.1.0} transforms to 
\begin{equation}
	\label{b.2.0}
	K(\zeta) = -\int^\zeta \left ( \frac{dW}{d\zeta} \right )^2 Z_\zeta^{-1} d\zeta, \qquad Z_\zeta \equiv {d Z \over d \zeta}.
\end{equation}
Analysis of the behavior near $\zeta=0$ yields
\begin{equation}
	\label{b.3}
	\left ( \frac{dW}{d\zeta} \right )^2 Z_\zeta^{-1} = E(\zeta) + O (\zeta),
\end{equation}
where
\begin{equation}
	\begin{split}
		\label{b.4}
		E(\zeta) &= 
		{1 \over \zeta^4} \biggl \{ \alpha^2 - \zeta^2 \left ( 2 a_2 \alpha^2
		+2 {\rm i} {\Gamma_0} \alpha  \right ) - 2 a_3 \alpha^2 \zeta^3 
		- \zeta^4 \left [2 \alpha^2 +{{\Gamma_0}}^2 - 4 {\rm i} a_2 {\Gamma_0} \alpha
		- (3 a_2^2 - 2 a_4 ) \alpha^2 \right ] \biggr \}.
	\end{split}
\end{equation}
Therefore, 
\eqref{b.2.0} implies
\begin{equation}
	\label{b.5}
	\overline{K}(\zeta)= - \int_0^\zeta  
	\left \{ \left ( \frac{dW}{d\zeta} \right )^2 Z_\zeta^{-1}  - E(\zeta) \right \} d\zeta
	+ {\bar k}_{-2} \zeta^{-3} + {\bar k}_{0} \zeta^{-1} - 2 a_3 \alpha^2 \log \zeta 
	+ {\bar k}_1 +  {\bar k}_2 \zeta \ ,   
\end{equation}
where
\begin{equation}
	\label{b.7}
	{\bar k}_{-2} = \tfrac13 \alpha^2 \ ,~{\bar k}_{-1}=0 \ , 
	\quad  {\bar k}_{0} = 
	-2 a_2 \alpha^2
	-2 {\rm i} {\Gamma_0} \alpha  
	\ , \quad {\bar k}_2 =  
	2 \alpha^2 + {\Gamma_0^2}  
	- 4 {\rm i} a_2 {\Gamma_0} \alpha
	- (3 a_2^2 - 2 a_4 ) \alpha^2 
\end{equation}
and the choice of integration constant in \eqref{b.2.0} is reflected
in the choice of ${\bar k}_1$. 
The appearance of $\log \zeta$ in the Laurent series 
in \eqref{b.5}
for ${\bar K} (\zeta) $ for $|\zeta| < 1$, and hence
in $K (1/\zeta)$  for $|\zeta| > 1$, is incompatible with 
\eqref{h.11} since the Laurent series of the left hand side
is single valued; this implies\footnote{It turns out
	any solution, at least in the parameter range investigated, 
	$a_{2j+1}=0$
	and hence $a_3=0$ is automatic.}
$a_3=0$, and therefore
\eqref{b.5} reduces to 
\begin{equation}
	\label{b.6}
	\overline{K}(\zeta) = - \int_0^\zeta  
	\left \{ \left ( \frac{dW}{d\zeta} \right )^2 Z_\zeta^{-1}  - E(\zeta) \right \} d\zeta
	+ {\bar k}_{-2} \zeta^{-3} + {\bar k}_{0} \zeta^{-1} 
	+ {\bar k}_1 +  {\bar k}_2 \zeta.
\end{equation}
It is important to note that, so far,
\eqref{b.2.0} has only been used to identify the first few coefficients 
${\bar k}_{-2}, {\bar k}_{0}, 
{\bar k}_1, {\bar k}_2$ of the 
the Laurent expansion of $\overline{K}(\zeta)$ at the origin (and hence
of the conjugate function $K(\zeta^{-1} )$ at infinity). 
Beyond these coefficients, the 
functional form of $\overline{K}(\zeta)$ in \eqref{b.2.0} 
is not used in what follows. Instead, 
as for the vortex problem, the function $\overline{K}(\zeta)$, and hence its conjugate
function, is determined through \eqref{h.11}.  
{\it A priori}, it would appear that there is no reason for the 
two expressions for
$\overline{K}(\zeta)$ in \eqref{h.11} and \eqref{b.6} 
to be consistent and, without further inspection, the bubble problem appears to be
overdetermined.
What is remarkable
is that there is consistency between these two functional representations
of $\overline{K}(\zeta)$ when a scalar constraint between three parameters
$\alpha, \Gamma_0$ and $\gamma$ is met. This consistency
manifests itself as a conformal map identity 
that is no less
valid for
the hollow vortex problem when the strain and shear components 
${\hat \alpha}, \beta$ and $\gamma$ satisfy 
the analogous constraint. 
This feature will be seen explicitly for the case
$\gamma=0$, presented in \S \ref{unified} where the resulting exact solutions are examined,
while the same has been proved for small $\gamma \ne 0$ in
the companion paper \cite{Valentin}. 
Numerical evidence in \S \ref{numerical}
suggest the same to be true beyond the small $\gamma$ regime.

\subsection{Alternative formulation of the bubble problem}

It is worth including a remark on an alternative formulation of the bubble problem that differs from
the formulation \eqref{b.1.1}.
From the expression for $W(Z(\zeta))$ as a
function of $\zeta$ in 
\eqref{b.2}, 
it follows from the chain rule that 
\begin{equation}
	\label{bub.1}
	\frac{dW}{dz} = \frac{\alpha \left (\zeta-\zeta^{-3} \right ) 
		{+}{\rm i} {\Gamma_0} \zeta^{-1}}{Z_\zeta}
	\ .
\end{equation}
The curvature $\kappa$ may be expressed in terms of
the conformal map on $|\zeta|=1$ as
\begin{equation}
	\label{bub.2}
	\kappa = \frac{1}{|y(\zeta)|^2} 
	\left (  1 + 2 \Re \left \{ \frac{\zeta y_{\zeta} (\zeta)}{y (\zeta)} 
	\right \} \right ),
\end{equation}
where $y(\zeta)$ is given by (\ref{yeqn}). With the earlier choice of 
length and velocity scales, the Bernoulli condition on the bubble
boundary, which is  
equivalent to \eqref{b.1.0.0}, 
implies that on $|\zeta|=1$:
\begin{equation}
	\label{bub.3}
	\frac{1}{|y(\zeta)|^2} \Bigl ( 1 + 2 \Re \left \{ \frac{\zeta y_{\zeta} (\zeta)}{
		y (\zeta)} \right \}  
	\Bigr )
	+ \left| \frac{dW}{dz} \right|^2 = \gamma,
\end{equation}
where ${dW}/{dz}$ is given by 
\eqref{bub.1}.
A proof of the existence of a solution to
\eqref{bub.3}, independent of
the formulation \eqref{b.1.1}, is given
in a companion paper \cite{Valentin} for small $\gamma$. 
This proof shows that 
in the formulation \eqref{b.1.1}, 
two different
expressions for $\overline{K} (\zeta)$ for the bubble problem, namely
\eqref{h.11}
and \eqref{b.6},
are automatically consistent when
a constraint between the strain rate $\alpha$, circulation ${\Gamma_0}$ and
Bernoulli constant $\gamma$ is satisfied.

\section{Unified representation for the vortex and bubble problems \label{unified}}

On introducing the parameters
\begin{align}
	\label{u.1}
	\lambda &= - {\hat \alpha} \ , \quad \sigma = - {\bar \beta} {+\gamma}. \quad 
	{\rm for~the~vortex~problem} \ , \\
	\lambda &= {\alpha^2 \over 3} \ , \quad \sigma = 2 {\rm i} \alpha {\Gamma_0},
	\quad 
	{\rm for~the~bubble~problem},
	\label{u.1.0}
\end{align}
then, concerning the Laurent series of $K (\zeta^{-1})$ 
in either case,
\eqref{h.10}
and \eqref{b.7} imply
\begin{equation}
	\label{u.2}
	k_{-2} = \lambda \ , \quad k_{0} = \sigma -6 \lambda {\bar a}_2.
\end{equation}
  {Identifying 
	$(\lambda, \sigma) $ with vortex and bubble 
	parameters in \eqref{u.1}, \eqref{u.1.0}, 
	from equations \eqref{h.3}, \eqref{h.4} for the vortex, 
	and equations \eqref{b.1.1}, \eqref{b.1.2}, \eqref{b.7} and
	\eqref{b.6} for the bubble imply that in either case,
	{\it Problem 3}, as stated earlier, is satisfied.}

Rewriting \eqref{h.11} in the form
\begin{equation}
	\label{u.3}
	\zeta y (\zeta) = \gamma {\overline{y}}(\zeta^{-1}) Z (\zeta) 
	+ K( \zeta^{-1} ) , 
\end{equation}
and plugging 
the series representations 
\eqref{h.5},\eqref{h.6} and \eqref{h.9} 
leads to a set of equations on
matching powers of $\zeta^j$ for $j \ge 0$.
With $k_{-2}, k_{-1}$ expressed in terms of $\lambda, \sigma$ in \eqref{u.2},
the $j=0$ equation is
\begin{equation}
	\label{h.12}
	k_2 = 1 + 
	\gamma - 
	\gamma \sum_{k=2}^\infty c_k {\bar a}_k 
	- \left \{ {\bar a}_4 \lambda + {\bar a}_2 (\sigma-6\lambda {\bar a}_2 )
	\right \}.
\end{equation}
Matching the coefficient of $\zeta$ in the Laurent series 
of \eqref{u.3} leads to 
\begin{equation}
	\label{h.13}
	k_1 = -\gamma \sum_{k=2}^\infty  
	c_{k+1} {\bar a}_k - \lambda {\bar a}_3.
\end{equation}
Equation \eqref{h.13} determines $k_1$ once $\{ a_j \}, \{c_j \}$ are determined.
More generally, for $j \ge 2$, 
matching of coefficients in
\eqref{u.3} leads to
\begin{equation}
	\label{h.17}
	a_j = \gamma c_j + \gamma \sum_{m=2}^\infty 
	c_{m+j} {\bar a}_m + \delta_{4,j} \lambda + 
	\delta_{2,j} \left ( \lambda {\bar a}_2
	+ [\sigma-6 \lambda {\bar a}_2 ] \right ).
\end{equation}
Equations \eqref{h.17} and \eqref{h.7} completely determine
$\left \{ a_j \right \}_{j=2}^\infty$, $\left \{ c_j \right \}_{j=2}^\infty$
for specified $\lambda, \sigma$ as has been rigorously proved in the
companion paper for sufficiently small nonzero $\gamma$ \cite{Valentin}. It also forms
the basis of the numerical calculations reported later. 

On the other hand, equating coefficients of   
$\zeta^{-j}$ for $ j \ge 1$ in 
the Laurent series 
of \eqref{u.3} leads to 
\begin{equation}
	\label{h.18}
	k_{j+2} = -\sum_{m=2}^{j+4} {\bar a}_m k_{j+2-m} - \gamma \sum_{m=2}^\infty
	{\bar a}_{m+j} c_m +\gamma {\bar a}_j.
\end{equation}
Equation \eqref{h.18} is a recurrence relation
to determine the set $\lbrace k_j  | j \ge 3 \rbrace$ since
$k_2,  \dots k_{-2}$, as well as the coefficients
$\left \{ a_j \right \}_{j=2}^\infty$ 
, $\left \{ c_j \right \}_{j=2}^\infty$ 
have been determined.

The important point is that,
in both problems, for specified $\lambda, \sigma$,
\eqref{u.3}
determines the Laurent series coefficient of both
$y(\zeta)$ -- and therefore $Z(\zeta)$ and $K(\zeta)$ --  implying
both the boundary shapes and the associated velocity fields are determined.

By equating the two different expressions for $k_2$ in 
\eqref{h.10}
and \eqref{h.12}
for the vortex problem,
the circulation parameter is found to be
\begin{equation}
	\label{eqomega}
	\omega  
	= -3 \lambda {\bar c}_4 + 3 \lambda {\bar c}_2^2 -\sigma {\bar c}_2 
	-1 -\gamma 
	+ 
	\gamma \sum_{k=2}^\infty c_k {\bar a}_k 
	+ \lambda {\bar a}_4 + {\bar a}_2 \left ( \sigma -3 \lambda {\bar c}_2 \right ).
\end{equation}

Moreover, the circulation is always real, as required for solutions to the vortex problem. This can be seen as follows \cite{Hstate}. Consider the equation  \eqref{h.4} and asymptotic expansion \eqref{h.3}. Using the fact that $\mathcal{Z}(s)$ is an arc-length parametrisation of $\partial \Omega$, we find that
\begin{align}
	-2\pi\omega =  \oint_{\partial \Omega} \xi(z) dz = \oint_{\partial \Omega} \left[\frac{d}{ds} \overline{\mathcal{Z}}(s) - i \gamma \overline{\mathcal{Z}}(s)\right] \frac{d \mathcal{Z}}{ds} ds =  |\partial \Omega| + 2\gamma |\Omega|  \label{e:CircReal}
\end{align}	
which expresses the circulation 
$-2\pi \omega$ in terms of the  perimeter $|\partial \Omega|$ and vortex area $|\Omega|$.\footnote{We are grateful to Ali Farsad for pointing out this argument.}


For the bubble problem, 
equating the two expressions for 
$k_2$ in 
\eqref{b.7}
and \eqref{h.12}  
leads to 
\begin{equation}
	\label{eqconst}
	2 \alpha^2 + {\Gamma_0}^2  +  
	4 {\rm i} {\bar a}_2 {\Gamma_0} \alpha - (3 {\bar a}_2^2 - 2 {\bar a}_4 ) \alpha^2  
	=
	1 +\gamma 
	- 
	\gamma \sum_{k=2}^\infty c_k {\bar a}_k 
	- {\bar a}_4 \lambda - {\bar a}_2 \left ( \sigma - 6 \lambda {\bar a}_2 \right ). 
\end{equation}
As has been proved in \cite{Valentin}, at least for small $\gamma$, 
when $\lambda \in \mathbb{R}$ and 
$\sigma \in {\rm i} \mathbb{R}$, 
as is the case for $\alpha, {\Gamma_0}$ real,
$c_{2j+1}=0=a_{2j+1}$, $c_{4j}, a_{4j}$ are real while
$c_{4j-2}, a_{4j-2}$ are imaginary.
This property implies 
\eqref{eqconst} is a real valued constraint
between the three real valued parameters 
${\Gamma_0}, \alpha$ and $\gamma$ that may be used
to determine straining rate $\alpha$ as a function of 
circulation ${\Gamma_0}$ and Bernoulli constant $\gamma$. 
An independent proof\cite{Valentin} for
the bubble problem shows that $\alpha$ is indeed a function of
$\gamma$ and ${\Gamma_0}$ at least for small $\gamma$.
Numerical evidence reported later
shows that this property holds beyond 
the small $\gamma$ limitation in the proof.

\subsection{Exact solution structure for $\gamma=0$}

When $\gamma =0$, then for $j \ge 2$,
\begin{equation}
	\label{e.1}
	a_j = \delta_{4,j} \lambda + \delta_{2,j} \left ( \lambda {\bar a}_2 + 
	\sigma-6\lambda {\bar a}_2 \right ).
\end{equation}
This implies that the only non-zero coefficients are $a_2$, and $a_4$. Substituting
$j=4$ in \eqref{e.1} gives 
$a_4 = \lambda$. Substituting $j=2$ in \eqref{e.1} 
gives an equation involving $a_2$ and its conjugate 
${\bar a}_2$. By taking the complex conjugate of that equation, and eliminating 
${\bar a}_2$ between the two expressions, gives 
\begin{equation}
	\label{e.2}
	a_2 = \frac{5 \lambda {\bar \sigma} - \sigma}{25 \lambda^2-1}.
\end{equation}
Since $Z_\zeta^{1/2} = y(\zeta) =1/\zeta + a_2 \zeta + a_4 \zeta$, it follows
that the derivative of the mapping function is the rational function
\begin{equation}
	\label{e.3} 
	Z_\zeta = \left ( {1 \over \zeta} +  
	\left ( \frac{5 \lambda {\bar \sigma} - \sigma}{25 \lambda^2-1} \right ) \zeta +
	\lambda \zeta^3 \right )^2.
\end{equation}
From \eqref{h.7}, the only nonzero values of $c_j$ are   
\begin{equation}
	\label{e.2.1}
	c_2 = 2 a_2, ~\ ~ c_4 = \tfrac23 \lambda + \tfrac13 a_2^2,~ \ ~
	c_6 = \tfrac25 a_2 \lambda, ~\ ~  c_8 = \tfrac17 \lambda^2.
\end{equation}
This can also be deduced by integrating \eqref{e.3}.  
A sufficient condition for the 
corresponding $Z (\zeta)$ 
to be univalent,  i.e., for boundary to be a simple closed curve,
has been proved \cite{Valentin}, which for the case $\sigma =0$, $\lambda > 0$ reduces to
the simple condition
\begin{equation}
	\label{equnivalent}
	1 > 2 \lambda + \lambda^2.
\end{equation}
{This condition is, however, not necessary: a plot of the shapes 
	reveal that they form a simple closed curve for $\lambda \in (0, 0.65)$,
	until pinch-off occurs
	for $\lambda \approx 0.66$.}

Instead of finding the Laurent series coefficients ${\bar k}_j$, 
we are using the equation
\begin{equation}
	\label{e.4} 
	\overline{K} (\zeta) = \frac{{\bar y}(\zeta^{-1})}{\zeta y (\zeta)}, 
\end{equation}
which follows from 
\eqref{u.3} upon setting $\gamma=0$,  
take its conjugate on $|\zeta|=1$ and find that
\begin{equation}
	\label{e.5}
	\overline{K} (\zeta) = \zeta^{-3} \left ( 
	\frac{\lambda + {\bar a}_2 \zeta^2 + \zeta^4}{1+a_2 \zeta^2 +\lambda \zeta^4} \right ).
\end{equation}

\begin{figure}[http]
	\centering
	\includegraphics[width=\textwidth]{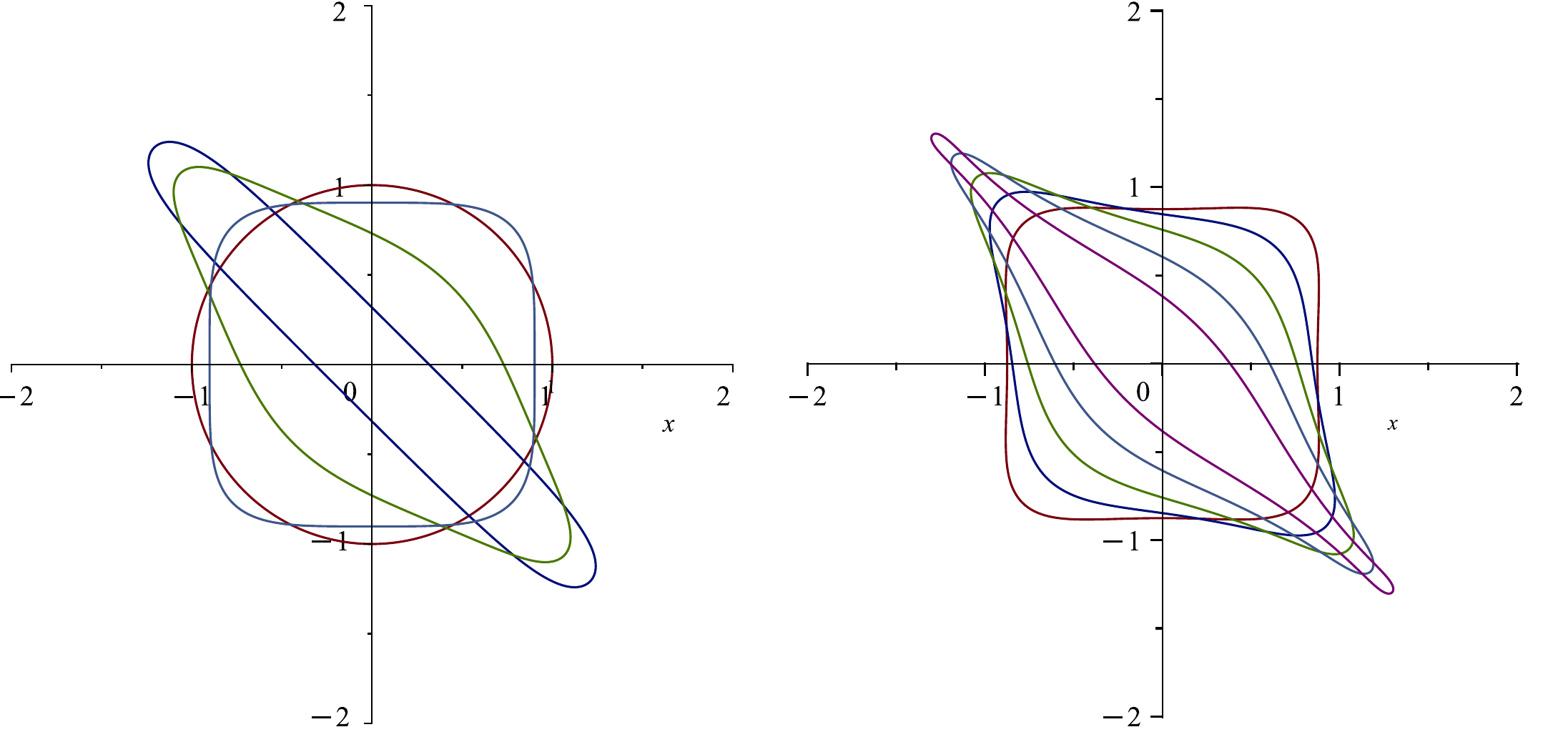}
	\caption{\textbf{Left:} Exact solution shapes with $\sigma, \lambda$ constrained by
		\eqref{b.11.1} for $\lambda = 0, 1/49, 6/49, 1/7$, respectively. As $\lambda$ increases 
		the shapes start off with a circle for $\lambda=0$, 
		then elongates with increasing $\lambda$ along $\Im z = - \Re z$ axis upto a point, before
		shortening and fattening into a a rhombus-like shape at $\lambda = 6/49$ until it 
		resembles a square with smoothed-out corners at $\lambda=1/7$ when $\sigma=0$. \textbf{Right:} Exact solution shapes for $\lambda = 0.18$,
		$-{\rm i} \sigma = 0, 0.01, 0.02, 0.03, 0.04$ (elongation increases with increasing $|\sigma|$).
		For $-{\rm i} \sigma \in (0.04, 0.05)$ the
		shapes pinch.}
	\label{figExactconstrained}
\end{figure}

If the parameters are constrained by the relation
\begin{equation}
	\label{b.11.1}
	\sigma = \pm{{\rm i}} (2-10 \lambda) \sqrt{\frac{3 \lambda (1-7 \lambda)}{1+25 \lambda}}   
	\ ,
\end{equation}
with $\lambda \le \tfrac17$, then the 
corresponding shapes 
for a sequence of
$\lambda$ values
are shown on the left in Fig \ref{figExactconstrained} with 
the branch in \eqref{b.11.1} chosen with $\Im \sigma \ge 0$. 
For the $\Im \sigma < 0$ branch in \eqref{b.11.1}, the corresponding shapes are similar, but
with elongation axis along $\Re z = \Im z$, instead of $\Re z = - \Im z$.
As will be seen in what follows, 
the constraint \eqref{b.11.1}
is necessary for the bubble, though not for the vortex. 
What is worth emphasizing here is that the free boundary shapes displayed in Figure \ref{figExactconstrained}, left, are identical for both, the Vortex problem and the Bubble problem upon converting the parameters $\lambda$ and $\sigma$ to the corresponding vortex ($\hat{\alpha}$ and $\bar{\beta}$) and bubble ($\alpha$ and $\Gamma_0$) parameters via \eqref{u.1}. This coincidence of free boundary shapes is solely due to the parameter constraint \eqref{b.11.1}, and it was proved in \cite{Valentin} that the coincidence of free boundary shapes persists even for $\gamma \neq 0$ sufficiently small. This is also numerically observed in Figure \ref{fig10:bubblestreamline}, wherein we also see that even though the boundary shapes are identical, the flows exterior to the vortex/bubble differ significantly.

More general shapes, 
that correspond to vortex but not bubble solutions and are 
without the constraint \eqref{b.11.1} are shown on the right in
Fig \ref{figExactconstrained}
for a sequence of $\sigma \in {\rm i} \mathbb{R}$ for fixed $\lambda$.
When $\Im \sigma < 0$ is negative, the shapes
are similar, except that the elongation axis is along $\Im z = \Re z$. When
$\Re \sigma \ne 0$, the symmetry about $ \Re z = \pm{\Im z}$ is broken and the shapes resemble
an inverted $S$ (similar to those shown on the right in
Figure \ref{fig1:lVortex}, although those correspond to
$\gamma \ne 0$).

\subsubsection{Exact solution for the vortex problem with no shear: $\gamma=0$}

With $\lambda=-{\hat \alpha}$, $\sigma=-{\bar \beta}$, then 
\begin{equation}
	\label{v.1}
	a_2 = \frac{5 \beta {\hat \alpha} +{\bar \beta}}{25 {\hat \alpha}^2 -1 }  
\end{equation}
and
\begin{equation}
	\label{v.2} 
	Z_\zeta = \left ( {1 \over \zeta} + 
	\frac{\left ( 5 \beta {\hat \alpha} +{\bar \beta}\right ) 
	}{25 {\hat \alpha}^2 -1 } \zeta  
	- {\hat \alpha} \zeta^3 \right )^2  
\end{equation}
\begin{equation}
	\label{v.3}
	\overline{K} (\zeta) = {1 \over \zeta^3} \left ( 
	\frac{-{\hat \alpha}
		+ {\bar a}_2 \zeta^2 + \zeta^4}{1+a_2 \zeta^2 -{\hat \alpha} \zeta^4} \right )
\end{equation}
with $a_2$ given by \eqref{v.1}.
The circulation $\omega$ in this case reduces to 
\begin{equation}
	\label{v.4}
	\omega  
	= -1 - {\hat \alpha}^2 - |a_2|^2. 
\end{equation}

\subsubsection{Exact solution for the bubble problem with zero Bernoulli constant: $\gamma=0$}
Using $\lambda=\tfrac13 \alpha^2$, $\sigma=2 {\rm i} {\Gamma_0} \alpha $ we obtain in
this case the only nonzero $a_j$ to be 
\begin{equation}
	\label{b.8}
	a_2 = - \frac{6{\rm  i} {\Gamma_0} \alpha}{5 \alpha^2 - 3}   
	\ , a_4 = \tfrac13 \alpha^2 \ .
\end{equation}
The constraint 
\eqref{eqconst}
reduces to 
\begin{equation}
	\label{b.11}
	{\Gamma_0^2} = \frac{ (3-5 \alpha^2 )^2 
		( 3-7 \alpha^2) }{9 (25 \alpha^2 +3 )  
	}
\end{equation}
requiring $7 \alpha^2 \le 3$. 
{Note that in the $\lambda-\sigma$ formulation, the constraint \eqref{b.11} 
	translates to \eqref{b.11.1}. }
The $\gamma=0$ exact solution
\eqref{v.2} 
in the 
bubble parameters translates to
\begin{equation}
	\label{b.12}
	Z_\zeta = \left ( \frac{1}{\zeta} 
	- \frac{6 {\rm i} {\Gamma_0} \alpha \zeta}{5 \alpha^2 - 3 } 
	+ \frac{\alpha^2}{3} \zeta^3 \right )^2,
\end{equation}
with 
\begin{equation}
	\label{b.13}
	\overline{K}(\zeta) ={1 \over \zeta^3} \left ( \frac{\tfrac13 \alpha^2 
		+ {\bar a}_2 \zeta^2 + \zeta^4}{1+a_2 \zeta^2 + \tfrac13 \alpha^2 \zeta^4}      
	\right ),
\end{equation}
where $a_2$ is given by  
\eqref{b.8} with
$\Gamma_0$  determined from \eqref{b.11}.
With $\Gamma_0$ constrained in this manner,
it may be directly checked that 
\begin{equation}
	\label{b.14}
	\frac{d}{d \zeta}{\overline{K}} (\zeta) = -\left ( \frac{dW}{d\zeta} \right )^2 Z_\zeta^{-1}  
\end{equation}
which implies consistency, for $\gamma=0$, between
the two different expressions for ${\overline{K}}$ from 
\eqref{h.11}
and \eqref{b.6}.

{This consistency has a wider implication. For the vortex problem
	$W(Z(\zeta))$ as given in 
	\eqref{b.2}
	has no physical significance.
	Yet,
	if we identify parameters in the vortex and bubble problems through the $\lambda, \sigma$ 
	correspondence,
	{ i.e.}, substitute
	$\alpha^2 = -3 {\hat \alpha}$, $2 {\rm i} {\Gamma_0} \alpha = 
	-{\bar \beta}+\gamma$ (note that $\gamma=0$ here) in
	\eqref{b.2} and 
	\eqref{b.11},  
	it ie easily checked that 
	\eqref{b.14} remains valid. Automatic satisfaction of  
	\eqref{b.14} is a conformal map identity since 
	$\overline{K} (\zeta)$ on the left side 
	hand side has already been determined through
	\eqref{e.4}. The important point is that this identity   
	transcends the
	bubble problem.
	
	A similar conformal map identity
	holds 
	for $\gamma \ne 0$, 
	as has been rigorously proved in the companion paper \cite{Valentin} 
	for small $\gamma$, even when 
	no exact solution is known to exist. Numerical calculations reported in 
	\S \ref{numerical} 
	suggest that the identity holds beyond the small $\gamma$ regime in the proof.

	\section{Numerical calculations for $\gamma \ne 0$.} 
	\label{numerical}
	
	This section presents some calculations of the solutions to both the vortex and the bubble problem
	when $\gamma \ne 0$.
	
	For both problems, a symmetry
	property inherent in the sequence relations 
	\eqref{h.7} and
	\eqref{h.17} is exploited.
	If
	$\left \{ a_2, a_3, a_4, a_5,  \cdots \right \}$ and 
	$\left \{ c_2, c_3, c_4, c_5, \cdots \right \}$ 
	satisfies
	\eqref{h.7} and \eqref{h.17}, then so does 
	$\left \{ a_2, -a_3, a_4, -a_5, \cdots \right \}$ and 
	$\left \{ c_2, -c_3, c_4, -c_5, \cdots \right \}$. Therefore, 
	when the solution is unique, a property proved to be true \cite{Valentin} 
	for solutions that continue from the $\gamma=0$ exact solution,
	then necessarily $a_{2j+1}, c_{2j+1} =0$ for
	all $j \ge 1$. 
	This implies that if $(x_p, y_p) $ is a point on the bubble or vortex boundary 
	corresponding to $\zeta = e^{{\rm i} \nu}$, then so is $(-x_p, -y_p)$ corresponding to
	$\zeta = - e^{i{\rm i}\nu}$, that is, there is an inversion symmetry that will be exploited in what follows.
	Whether or not there are
	solutions without inversion symmetry remains an open question at this time.    
	
	There is another symmetry inherent in 
	equations \eqref{h.7} and \eqref{h.17} when
	$\sigma $ is imaginary. If
	the sequences
	$\left \{ a_2, a_3, a_4, a_5,  \cdots \right \}$ and 
	$\left \{ c_2, c_3, c_4, c_5, \cdots \right \}$ satisfy these equations,
	so do
	$\left \{ {\bar a}_2, -{\bar a}_3, {\bar a}_4, -{\bar a}_5,  \cdots \right \}$ and 
	$\left \{ {\bar c}_2, -{\bar c}_3, {\bar c}_4, -{\bar c}_5, \cdots \right \}$.
	Therefore, when the solution is unique as has been proved\cite{Valentin} for 
	small $\gamma$, and confirmed numerically when $\gamma$ is not small,
	then $\left \{ a_{4j}, c_{4j} \right \}_{j=1}^\infty$ is real and 
	$\left \{ a_{4j-2}, c_{4j-2} \right \}_{j=1}^\infty$ imaginary.
	This implies that $e^{-{\rm i} \pi/4} Z$ has a power series involving only odd powers
	$e^{{\rm i} \pi/4} \zeta$ with all series coefficients real.  
	This implies that the shapes are symmetric about
	$\Re z = \pm{\Im z}$.
	Therefore, with this additional symmetry, if
	$(x_p,y_p)$ is a point on the boundary, so is $(y_p, x_p)$, $(-y_p, x_p)$, besides
	$(-x_p,-y_p)$. However, for the  vortex problem, when $\beta - \gamma$ is not
	imaginary, there are no symmetries about $\Im z = \pm{\Re z}$
	and only
	the inversion symmetry pertains.
	For the bubble, in all cases, besides inversion symmetry, 
	the shapes have symmetry about
	$\Re z = \pm{\Im z}$ 
	since
	$\sigma = 2 {\rm i} {\Gamma_0} \alpha $ is imaginary. 
	
	For given $\lambda, \sigma$, the coefficients
	$a_j, c_j$ for $j \ge 2$ are determined by using
	\eqref{h.7} and \eqref{h.17};  note that  
	$k_{-2}, k_0$ are related to $\lambda, \sigma$ through \eqref{u.2}.
	The Laurent series coefficents $ k_2, k_3, \dots$ 
	emerging from 
	\eqref{h.13} and \eqref{h.18} are used  to determine
	$\xi (Z (\zeta)) = {\rm i} \overline{K} (\zeta)$ and hence the fluid velocity in the domain $\Omega$.
	
	When 
	$\lambda, \sigma, \gamma$ are constrained by
	\eqref{eqconst} with the correspondence
	$\alpha^2 = 3 \lambda$, $2 {\rm i} \Gamma_0 \alpha = \sigma $,
	for the 
	$\overline{K} (\zeta)$ determined 
	through its Laurent series coefficients 
	${\bar k}_{-2}, {\bar k}_{0}, {\bar k}_2, \cdots$, 
	the conformal map identity, namely,
	\begin{equation}
		\label{eq:conformal}
		\frac{d}{d \zeta} \overline K (\zeta) = - \frac{1}{Z_\zeta} \left ( \frac{dW}{d\zeta} \right )^2,
	\end{equation} 
	where 
	\begin{equation}
		\label{eq:dWdzetageneral}
		\frac{dW}{d\zeta} = 
		\sqrt{\frac{3}{\lambda}}
		\left \{ \frac{\sigma}{6 \zeta} + \lambda 
		\left ( \zeta - \zeta^{-3} \right ) 
		\right \} \ ,
	\end{equation}
	was checked numerically for both the vortex and bubble problems
	by evaluating both left and right hand sides of \eqref{eq:conformal}
	at a discrete set of uniformly spaced points on $|\zeta|=1$. To within small numerical errors, 
	\eqref{eq:conformal}
	remained valid 
	beyond the small $\gamma$ limitation of the rigorous proof \cite{Valentin}.
	
	\subsection{Numerical solutions of the vortex problem}
	
	For given third and first order straining components 
	${\hat \alpha} $, $\beta$ and shear $\gamma$, 
	i.e., specifying $\lambda, \sigma, \gamma$, 
	the infinite 
	sequence
	$ \left \{ a_{2m} \right \}_{m=1}^{N}$ is truncated; that is,
	$a_{j} =0$ outside this range.
	For specified $ \left \{ a_{2m} \right \}_{m=1}^{N}$, 
	the set $ \left \{c_{2m} \right \}_{j=1}^N$ is determined through 
	\eqref{h.7}. Equation
	\eqref{h.17} is satisfied
	for all even indices $j$ from 2 to $2N$, noting that 
	for $j$ odd, \eqref{h.17} is automatically satisfied from the assumed symmetry.
	This results in a system of $N$ complex equations for $N$ complex unknowns. Treating the
	real and imaginary parts of $a_{2m}$ as unknowns, and taking real and imaginary
	parts of \eqref{h.17}, the resulting system of $2N$ real algebraic equations 
	for $2N$ unknowns is solved via Newton iteration. The $\gamma=0$ exact solution
	provides a good initial guess for small $\gamma$ which is then continued both for
	positive and negative $\gamma$. Univalence of the conformal map $Z(\zeta)$ is checked
	by plotting the resulting shapes.
	Since odd coefficients $a_{2j+1}, c_{2j+1}$ are zero, it is clear that 
	from \eqref{h.13} that
	$k_1 =0$ therefore from \eqref{h.10}, there is no drift,  i.e., $\delta=0$.
	The circulation term $\omega$ is determined from 
	\eqref{eqomega},  and always real, 
	as shown in equation \eqref{e:CircReal}.

	{Streamlines are generated using
		the differential equation $dz/dt = u+{\rm i} v$. It is found more convenient
		to determine streamline images in the $\zeta$ plane since
		$\xi(Z(\zeta)) = {\rm i} \overline{K} (\zeta)$ is known in $|\zeta| \le 1$
		through the Laurent series coefficients of $K(\zeta^{-1})$, 
		determined from \eqref{h.12}, \eqref{h.13} and 
		\eqref{h.18}.
		The image of the streamline in the $\zeta$-plane 
		is generated by numerically solving the ODE
		\begin{equation}
			\label{eqdiffstreamv}
			\frac{d\zeta}{dt} = -{{\rm i} \over Z_\zeta} \left \{ \gamma Z (\zeta) + 
			K \left ({\bar \zeta} \right )  \right \}
		\end{equation}
		with initial condition determining the particular streamline. This is translated to
		the $z$ plane 
		through 
		\eqref{h.5}.
		Note that 
		${\bar \zeta} \ne \zeta^{-1}$ 
		for $|\zeta|  < 1$ unlike on
		$|\zeta|=1$.
		In the figures below, shapes are presented for fixed $\lambda, \beta$ for
		a range of $\gamma$ and  the corresponding areas determined from
		the relation
		\begin{equation}
			\label{eqarea}
			A = \pi \left ( 1 - \sum_{n=2}^\infty (n-1) |c_n|^2 \right ).
		\end{equation}
		The circulation parameter $\omega$ 
		as a function of shear $\gamma$ for fixed straining components
		${\hat \alpha}$ and $\beta$ is also displayed.
		
		
		
		\begin{figure}[h]
			\centering
			\includegraphics[width=\textwidth]{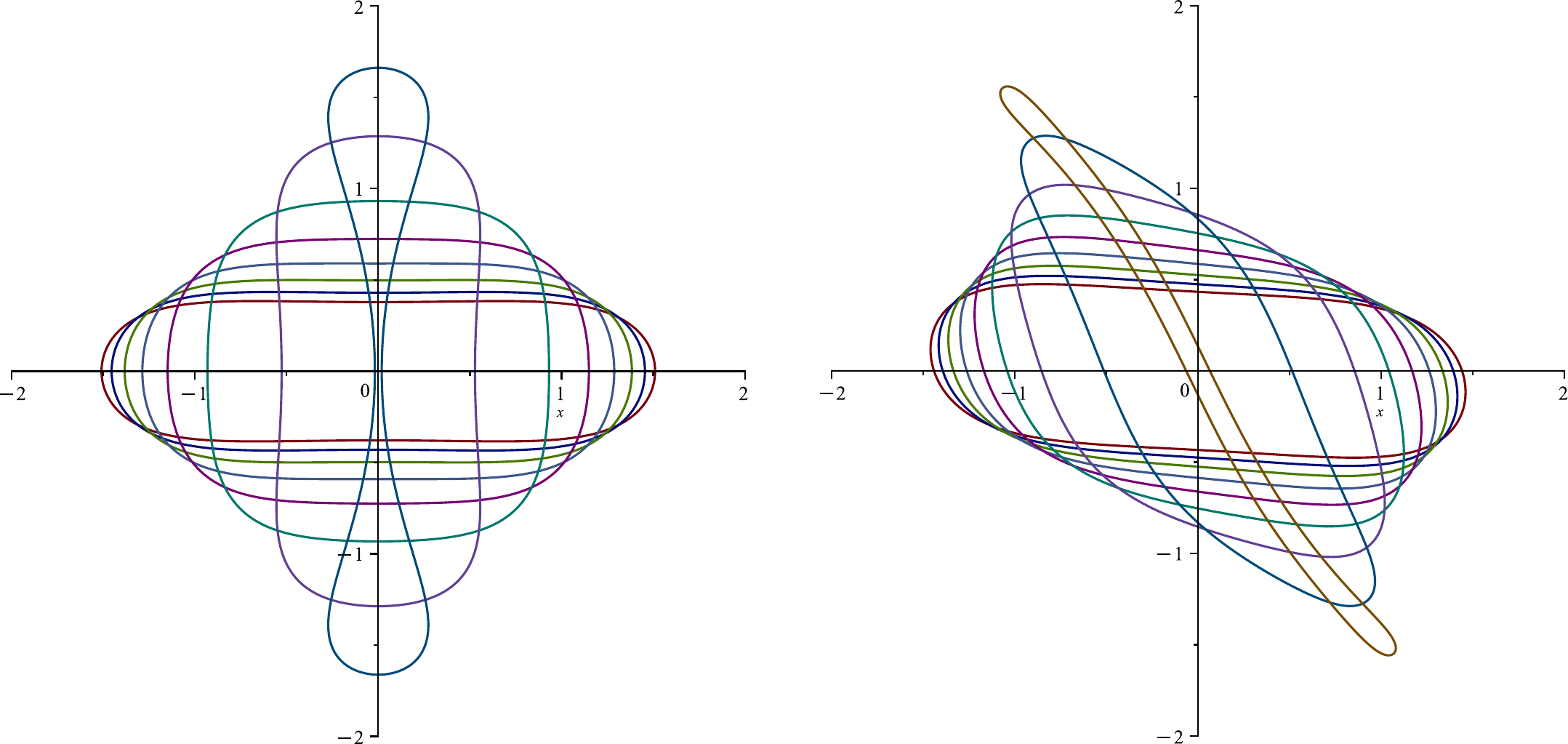}
			\caption{Vortex shapes.\textbf{ Left:} strain components ${\hat \alpha} = -1/10$, 
				$\beta = 0$, for shear $\gamma = -1,-0.8,-0.6,-0.40,-0.20,0,0.20,0.32$
				with corresponding areas 
				2.1187, 2.2982, 2.5044, 2.7351, 2.9678, 3.09966, 2.6326, 0.9274,
				respectively. The $\gamma=-1$ shape corresponds to the most flattened
				elliptical shape; the shapes gets progressively elongated vertically
				with increasing $\gamma$ until it pinches for $\gamma \in (0.32, 0.33)$. \textbf{Right:} Strain components ${\hat \alpha} = -1/10$, 
				$\beta = -1/10+i/10$, for shear $\gamma = -1,-0.85,-0.7,-0.55,
				-0.40,-0.25,-0.10,0.05,0.13$
				with corresponding areas 
				2.2890, 2.4174, 2.5544, 2.6951, 2.8265, 2.9135, 2.8513, 2.1881, 0.4523
				, respectively. The $\gamma=-1$ shape roughly resembles an ellipse and gets progressively deformed
				with increasing $\gamma$ until it pinches for $\gamma \in (0.13,0.14)$.}
			\label{fig1:lVortex}
		\end{figure}
		
		\begin{figure}[h]
			\centering
			\includegraphics[width=\textwidth]{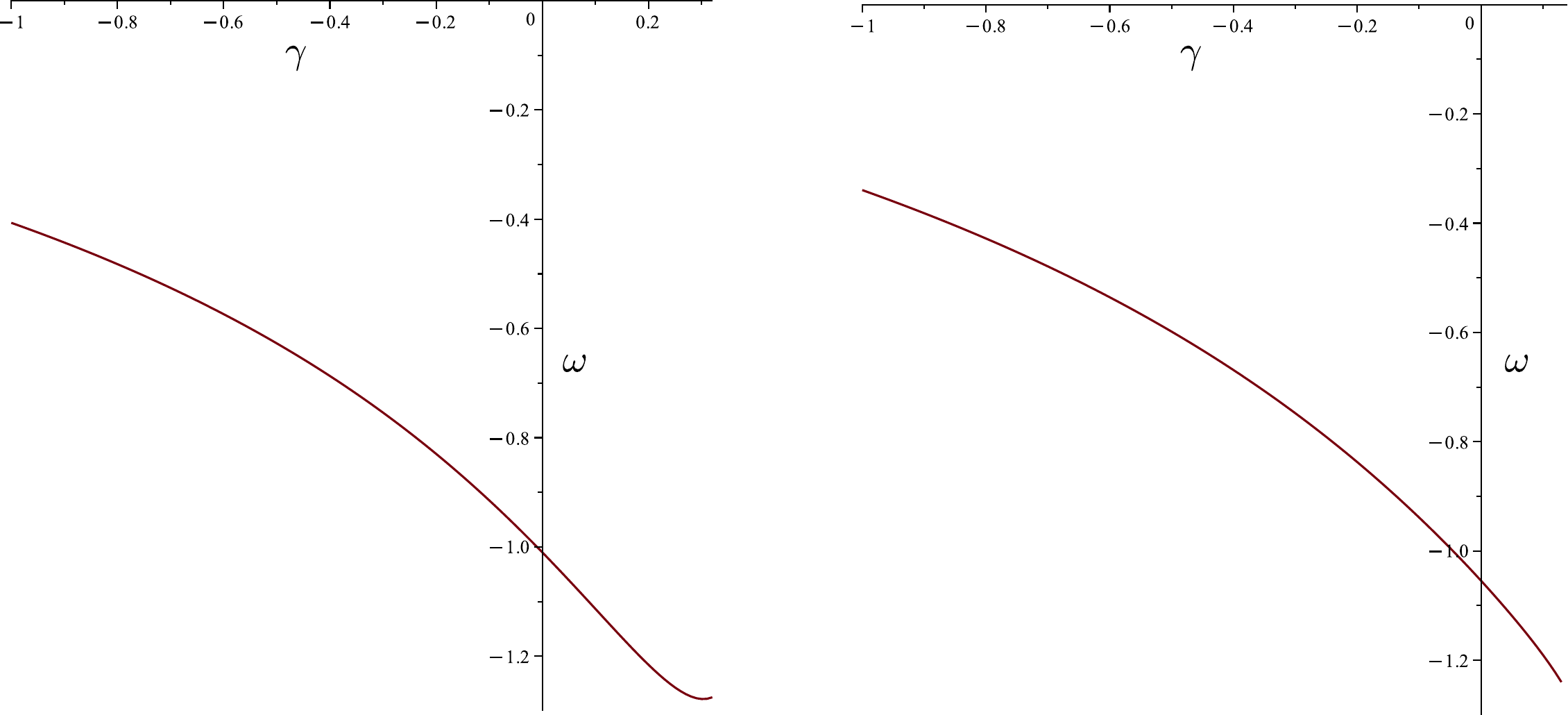}
			\caption{Circulation parameter $\omega$ against shear $\gamma$ for 
				given strain 
				components. \textbf{Left:} ${\hat \alpha} = -1/10$, 
				$\beta = 0$. \textbf{Right:} ${\hat \alpha} = -1/10$, 
				$\beta = -1/10+i/10$.}
			\label{fig2:circulationagainstshearasymmetricstrainplusshear}
		\end{figure}

		\FloatBarrier
		\subsection{Numerical solutions of the bubble problem}
		
		
		In this section solutions for the bubble shapes have been calculated in two ways.
		
		For specified circulation ${\Gamma_0}$ and Bernoulli constant $\gamma$, 
		as for the vortex case, 
		the infinite 
		sequence
		$ \left \{ a_{2m} \right \}_{j=1}^{N}$ is truncated
		to $N$ terms 
		and $ \left \{c_{2m} \right \}_{j=1}^N$ is determined from 
		\eqref{h.7}.
		As before,  \eqref{h.17} is satisfied for even indices $j$ from 2 to $2N$
		(with odd indexed equations automatically satisfied due to the assumed symmetry)
		and this set of equations is supplemented with the constraint 
		\eqref{eqconst}. This gives a set of $2N+1$ real equations for the $2N+1$ unknowns
		$\alpha, \left \{ \Re a_{2j}, \Im a_{2j} \right \}_{j=1}^N$ 
		which is satisfied via Newton iteration. The $\gamma=0$ exact
		solutions provide a good initial guess for small $\gamma$ and 
		continuation in $\gamma$ was carried out for both positive and negative $\gamma$.   
		It is to be noted that
		$\alpha, {\Gamma_0}$ determines $\lambda, \sigma$ that appear in
		\eqref{h.17}, 
		The shapes for given circulation ${\Gamma_0}$ for a range of $\gamma$ are shown in 
		Figure \ref{fig5:bubblep0}. The strain-rate parameter $\alpha$ is also plotted as a function of
		$\gamma$ for two fixed values of ${\Gamma_0} = 0, 0.1$
		in Figure \ref{fig6:alphabubblep0}. As already mentioned, 
		the case of zero circulation
		${\Gamma_0}=0$ has been calculated before by Vanden Broeck \& Keller \cite{Vanden}  using
		a different approach.  
		When $\Gamma_0 < 0$ was chosen negative, the shapes resemble the ones for
		$\Gamma_0 > 0$ for the same Bernoulli constant
		except that the elongation axis is
		along $ \Im z = \Re z$, instead of $\Im z = - \Re z$, seen in
		Figure 
		\ref{fig5:bubblep0}, right,
		for $\Gamma_0 = 0.1$.
		
		A second numerical method relied on
		satisfying 
		\eqref{bub.3}
		at $2N-1$  
		uniformly spaced out points on the unit circle $|\zeta|=1$
		with a truncated representation of $y$ by 
		\begin{equation}
			\label{eqyt}
			y_t (\zeta) = {1 \over \zeta} + \sum_{m=2}^{N} a_m \zeta^{m-1}
		\end{equation}
		and with $\left \{\Re a_j \right \}_{j=2}^N , 
		\left \{\Im a_j \right \}_{j=2}^N $ along with $\alpha$ taken as the unknowns.
		Newton iteration was employed with the exact solution for $\gamma=0$ again providing a good initial
		guess for the solution
		for small $\gamma$. It is important to note that, in applying this method, it was not assumed that 
		$a_{2j+1} =0$, although the solutions reflected this property.
		Furthermore, the solution also reflected the property that for any $ j \ge 1$,
		$a_{4j} $ is real, while
		$a_{4j-2}$ is imaginary.
		
		Besides checking for programming errors, 
		the agreement between these two independent methods for
		finding  
		solutions, to within very small numerical errors,
		validates the consistency between two different expressions of 
		$\overline{ K} (\zeta)$ since the first approach
		produces a solution for a bubble only if the equations are consistent.
		
		Streamlines for the case of the bubble are
		generated by the differential equation $dz/dt = u+{\rm i} v = \overline{dW/dz} $:
		the  
		image of the streamline in the $\zeta$-plane 
		is generated through 
		\begin{equation}
			\label{eqdiffstreamv}
			\frac{d\zeta}{dt} = \frac{1}{Z_\zeta} \left \{ 
			\overline{ \frac{W_\zeta}{ Z_\zeta} }  \right \}  
			\ , 
		\end{equation}
		where $W(Z(\zeta) )$ is given by 
		\eqref{b.2}. 
		Different initial conditions give different streamlines. This is translated to
		the $z$ plane through $z=Z(\zeta) $ calculated 
		through 
		\eqref{h.5}.
	}
	
	%
	%
	
	Because of the relation between the bubble and vortex problem, the shapes that
	have been shown for the bubbles are exactly the same shape for the hollow vortex
	provided  the scaled bubble circulation and Bernoulli 
	parameters $\Gamma_0$, $\gamma$ and corresponding
	strain $\alpha$ to hollow vortex parameters are 
	associated in the following manner: 
	${\hat \alpha} = - \alpha^2/3$, 
	$-{\bar \beta} + \gamma = 2 {\rm i} \alpha \Gamma_0$ with $\gamma$ now interpreted to be the
	shear rate.
	It was verified numerically 
	that this is indeed the case.    
	
	For a particular case, Figure \ref{fig10:bubblestreamline}
	shows the streamline distributions for both the vortex problem (left) and the bubble
	problem (right).
	%
	The  streamline patterns are quite distinct even though the boundary shapes are identical;
	this is to be expected since the values of parameters are 
	chosen so that the mathematical parameters $\lambda, \sigma, \gamma$ match exactly.
	
	\begin{figure}[h]
		\centering
		\includegraphics[width=\textwidth]{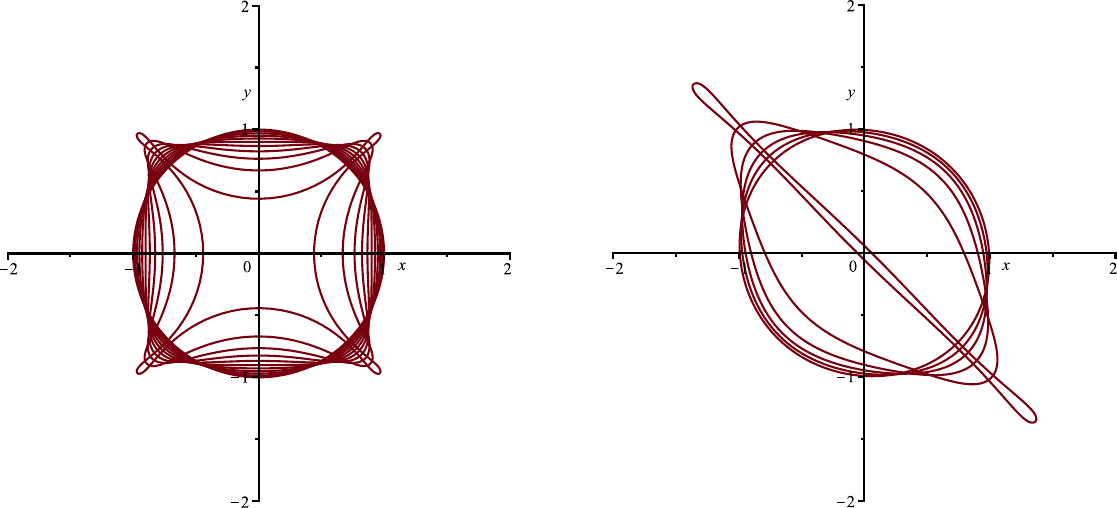}
		\caption{\textbf{Left:} Bubble shapes with no circulation 
			for $\gamma = -0.99,-0.79,-0.59,-0.39, -0.19, 0.01, 0.21, 0.41, 0.81, 1$
			with corresponding straining rates 
			$\alpha =
			0.070672, $ $0.32009,$ $ 0.44140, $ $0.53038,  $ $0.60067,$ $0.65719, 
			$ $0.70184,$ $0.73466,$ $0.75362,$ $0.75159,$ 
			$0.68940$ 
			and areas
			3.14159, 3.1395, 3.1325, 3.1184, 3.0938, 3.0534, 2.9878, 2.8797, 2.6911, 2.3126,
			1.0670.
			As $\gamma$ increases from -1 to 1.00, 
			the shapes deviate more and more from a circle and approaches a four-fold pinching 
			corner for $\gamma \in (1, 1.01)$.
			At $\gamma=-1$,
			we have a cicle of unit radius in a quiescent flow where
			$\alpha=0$. \textbf{Right:} Bubble shapes for scaled circulation $\Gamma_0 = 0.1 $
			for Bernoulli constants $\gamma = -0.9,-0.7,-0.5,-0.3$ and $0.16$
			with corresonding straining rates $\alpha \approx  
			0.20923, 0.36993, 0.47169, 0.54431, 0.58527,
			0.52444$ and areas
			3.1383, 3.1229, 3.0862, 
			2.9909,  
			0.2351891413636648, respectively. 
			As $\gamma$ increases from -1 to 0.16, 
			the shapes get more elongated from an initial circular shape, 
			until the shapes pinches for $\gamma \in (0.16, 0.17)$.}
		\label{fig5:bubblep0}
	\end{figure}

	\begin{figure}[h!]
		\centering
		\includegraphics[width=\textwidth]{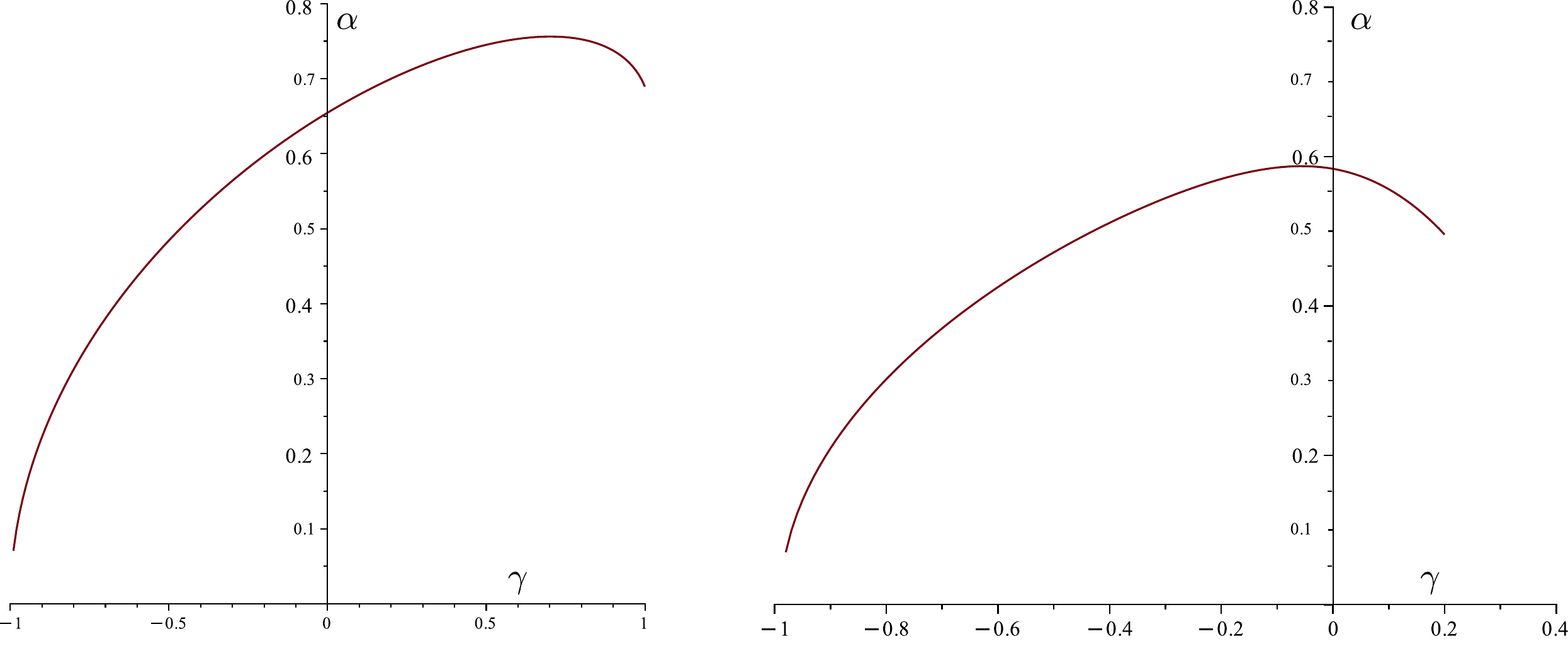}
		\caption{Strain rate $\alpha$ vs $\gamma$. \textbf{Left:} $\Gamma_0=0$, \textbf{Right:} $\Gamma_0 = 0.1 $} 
		\label{fig6:alphabubblep0}
	\end{figure}

	\begin{figure}[h!]
		\centering
		\includegraphics[width=\textwidth]{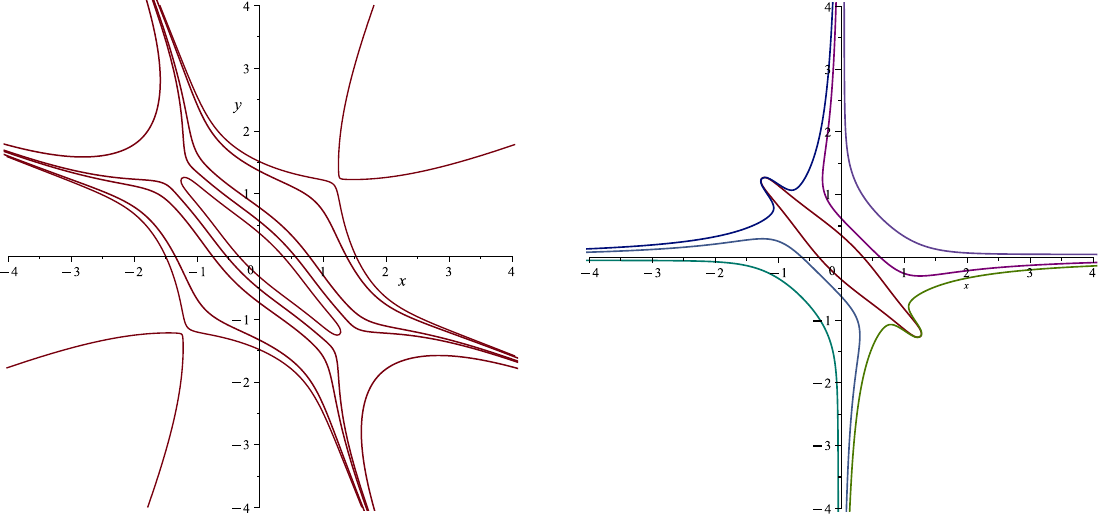}
		\caption{\textbf{Left (vortex):} Streamlines for ${\hat \alpha} = -0.103139$, $\beta_0=0.1 + 0.111251 {\rm i} $, 
			$\gamma=0.1$. The
			corresponding vortex area $\approx$ 1.2296.
			Note four stagnation points at $z \approx \pm{\left ( 1.35241- 1.35241 {\rm i} \right )} $, 
			$z \approx \pm{\left ( 1.23992 +1.23992 {\rm i} \right )}$ 
			on lines of symmetry $\Re z =\pm{\Im z}$ that arise since $\sigma={\bar \beta}-\gamma$ is
			imaginary. \textbf{Right (bubble):}  Streamlines for $\Gamma_0=0.1$, $\gamma=0.1$. The
			corresponding bubble area is 1.2296 and strain $\alpha=0.55625$.
			Note four stagnation points occur on the bubble boundary corresponding to
			$\zeta = \pm{e^{-{\rm i} \nu_s}}, \pm{i e^{-{\rm i} \nu_s}}$, where $\nu_s = \frac12 \arcsin \left ( 
			\frac{\Gamma_0}{2 \alpha} \right ) $.
			Note that for large $z$, the flow moves from $\mathrm{Im}z$ axis towards real $z$ axis, as
			expected from $dW/dz \sim \alpha z $ for $\alpha > 0$.
		}
		\label{fig10:bubblestreamline}
	\end{figure}

	\section{Discussion}

	Although this aspect is not explored here,
	it is expected that there exists a hierarchy of mathematical problems (Problem 4, Problem 5, etc.) for
	which there is also a pair of physically-distinct problems whose solutions fall within the set of solutions
	of each mathematical problem.
	Just as this study has been motivated by considering the steadily-rotating extensions
	of the $N=4$ solutions of Crowdy \& Roenby \cite{Roenby} shown in Figure \ref{CRFig},
	these higher-order problems will be relevant to finding steadily-rotating 
	analogues of the Crowdy-Roemby solutions for higher powers of $N$ where the inner problem near the central
	hollow vortex will need to match to a higher-order irrotational straining flow induced by the corotating  $N$-polygonal
	satellite point vortex array.
	And it is likely that, for the analogous capillary problem in each case, at least one constraint between parameters will
	be required
	for the solutions to be physically admissible.
	Whether it remains a {\em single} constraint as the order of the singularity of $\xi(z)$
	at infinity increases remains to be determined. The evidence here points in
	that direction, however.
	
	Unlike {\it Problem 2}, where the scalar constraint leads to a 
	single one-dimensional curve, $\gamma=\gamma(\beta)$, for {\it Problem 3}, 
	we have a single two-dimensional \textit{surface}, $\gamma = \gamma(\alpha,\Gamma_0)$.
	With higher order straining flow, the dimension of the manifold resulting from the scalar constraint
	is expected to further increase.

	\begin{itemize}
		\item  The codes used for the computations in this paper are available
		from the authors on request.
		\item All authors contributed to the contents of this paper.
		\item We declare that we have no competing interests.
		\item ST acknowledges support from Simons
		Foundation International through the 
		travel Grant (SF1-MPS-TSM00013918)
	\end{itemize}

	
	\vskip2pc
	
	\bibliographystyle{plain}
	\bibliography{bibliography_royal_society}

\end{document}